\nonstopmode % The compilation does not stop when encountering error. Usefull when compiling with Makefiles
\documentclass[%
reprint,
 amsmath,amssymb,
 aps,
]{revtex4-1}

\usepackage{graphicx}% Include figure files
\usepackage{dcolumn}% Align table columns on decimal point
\usepackage{bm}% bold math
\usepackage[utf8]{inputenc}
\newcommand{\bra}[1]{\langle #1|}
\newcommand{\ket}[1]{|#1\rangle}

\newcommand{\avg}[1]{\langle#1\rangle} % for average

\newcommand{\op}[1]{\hat{#1}} % Put here the command to format operators.
\newcommand{\ex}[1]{\mathrm{e}^{#1}} % Exponential
\newcommand{\abs}[1]{\left|#1\right|}
\newcommand{\dint}[4]{\int\limits_{#3}^{#4} \!#1\,\mathrm{d}#2}

\begin{document}

\preprint{APS/123-QED}

\title{Stimulated Raman adiabatic passage in a three-level superconducting circuit}% Force line breaks with \\
%\thanks{A footnote to the article title}%

%\author{Ann Author}
% \altaffiliation[Also at ]{Physics Department, XYZ University.}%Lines break automatically or can be forced with \\
%\author{Second Author}%
% \email{Second.Author@institution.edu}
%\affiliation{%
% Authors' institution and/or address\\
% This line break forced with \textbackslash\textbackslash}%

\author{K.~S. Kumar}
\affiliation{Low Temperature Laboratory, Department of Applied Physics, Aalto University School of Science, P.O. Box 15100, FI-00076 AALTO, Finland}
\author{A. Veps\"al\"ainen}
\affiliation{Low Temperature Laboratory, Department of Applied Physics, Aalto University School of Science, P.O. Box 15100, FI-00076 AALTO, Finland}
\author{S. Danilin}
\affiliation{Low Temperature Laboratory, Department of Applied Physics, Aalto University School of Science, P.O. Box 15100, FI-00076 AALTO, Finland}
\author{G.~S. Paraoanu}
\affiliation{Low Temperature Laboratory, Department of Applied Physics, Aalto University School of Science, P.O. Box 15100, FI-00076 AALTO, Finland}

%\collaboration{MUSO Collaboration}%\noaffiliation

%\author{Charlie Author}
% \homepage{http://www.Second.institution.edu/~Charlie.Author}
%\affiliation{
% Second institution and/or address\\
% This line break forced% with \\
%}%
%\affiliation{
% Third institution, the second for Charlie Author
%}%
%\author{Delta Author}
%\affiliation{%
% Authors' institution and/or address\\
% This line break forced with \textbackslash\textbackslash
%}%

%\collaboration{CLEO Collaboration}%\noaffiliation

\date{5 August 2015}% It is always \today, today,
             %  but any date may be explicitly specified

\begin{abstract}
\bf{
The adiabatic manipulation of quantum states is a powerful technique that has opened up new directions in quantum engineering, enabling tests of fundamental concepts such as the Berry phase %\cite{berrywallraff}
and its nonabelian generalization, %\cite{nonabelianwallraff},
the observation of topological transitions, %\cite{topologicaltransitions}
and holds the promise of alternative models of quantum computation. %\cite{annealing,dwave,troyer,fastholonomic,holonomic,topologicalQCbook}.
Here we benchmark the stimulated Raman adiabatic passage %(STIRAP)
process %\cite{reviewSTIRAP,another_reviewSTIRAP}
for circuit quantum electrodynamics, by using the first three levels of a transmon qubit. We demonstrate  a population transfer efficiency above $80 \%$ between the ground state and the second excited state using two adiabatic Gaussian-shaped control microwave pulses coupled to the first and second transition. The advantage of this techniques is robustness against errors in the timing of the control pulses. By doing quantum tomography at successive moments during the Raman pulses, we investigate the transfer of the population in time-domain. %Using this method, we realize a high fidelity quantum gate which can produce arbitrary superposition states between the ground state and the second excited state.
We also show that this protocol can be reversed by applying a third adiabatic pulse on the first transition. Furthermore, we demonstrate a hybrid adiabatic-nonadiabatic gate using a fast pulse followed by the adiabatic Raman sequence, and we study experimentally the case of a quasi-degenerate intermediate level.}
%\begin{description}
%\item[Usage]
%Secondary publications and information retrieval purposes.
%\item[PACS numbers]
%May be entered using the \verb+\pacs{#1}+ command.
%\item[Structure]
%You may use the \texttt{description} environment to structure your abstract;
%use the optional argument of the \verb+\item+ command to give the category of each item.
%\end{description}
\end{abstract}

\pacs{Valid PACS appear here}% PACS, the Physics and Astronomy
                             % Classification Scheme.
%\keywords{Suggested keywords}%Use showkeys class option if keyword
                              %display desired
\maketitle

%\tableofcontents
%\section*{Introduction}

The precise control and manipulation of the states of multilevel quantum systems is a key requirement in quantum information processing.
Using multilevel systems as the basic components of quantum processors instead of the standard two-level qubit brings along the benefit of an extended Hilbert space, thus ultimately reducing the number of circuit elements required to perform a computational task \cite{Lanyon2009}. %\cite{qutrit_logic_luo, Lanyon2009}.
Superconducting qubits can be operated as multilevel systems, and to date fundamental effects related to three-level artificial atoms irradiated with continuous microwave fields - for example the Autler-Townes effect \cite{AT_us,AT_wallraff}, coherent population trapping \cite{coherent_population_trapping}, and electromagnetically-induced transparency \cite{EIT_abdumalikov} - have been reported. An additional powerful concept is the adiabatic control of quantum systems, which in recent times led to new directions in quantum engineering, enabling tests of fundamental concepts such as Berry phase \cite{berrywallraff}
and its nonabelian generalization \cite{nonabelianwallraff},
the observation of topological transitions \cite{topologicaltransitions}, and holds the promise of alternative models of quantum computation \cite{annealing,dwave,troyer,fastholonomic,holonomic,topologicalQCbook}.

In this paper we employ a superconducting circuit irradiated with two microwave fields to demonstrate a fundamental quantum-mechanical process called stimulated Raman adiabatic passage (STIRAP) \cite{reviewSTIRAP,another_reviewSTIRAP}: the transfer of population between two states $\ket{0}$ and $\ket{2}$ via an intermediate state $\ket{1}$ using two control microwave fields, with zero occupation of the intermediate state at any time. The absence of population in the intermediate state is ensured by a two-photon destructive interference on the intermediate level which creates a dark state, while the adiabatic manipulation of the amplitudes of the control fields rotates the dark state from the initial state $\ket{0}$ to the target state $\ket{2}$. The STIRAP technique was first demonstrated with sodium dimer molecular beams \cite{stirapfirst} and applied since then in many contexts in atomic physics. It has remained a topic of active research \cite{experimental_atom_du}, and in recent years unexpected connections and analogies have been found.
Recently there have been proposals on designing $\Lambda$ systems using superconducting circuits with similar energy structure as atoms \cite{stirap_cooper_pair_box_falci,lambda_quantronium_falci}. However, due to the present pre-eminence of the transmon as the building block of the future superconducting quantum processors, it would be most relevant to demonstrate this technique on a typical device of this kind. The transmon \cite{transmon_PRA2007} is a capacitively-shunted Cooper pair box strongly coupled to an electromagnetic transmission line resonator \cite{cqed_strong_coupling_wallraff}, which allows the quantum non-demolition measurement of the state of the transmon by probing the resonator's resonance frequency with microwaves \cite{dispersive_readout_bianchetti}. Here we aim at benchmarking the STIRAP protocol under realistic conditions - with a transmon that has finite decoherence times associated to the excited states.

%In circuit quantum electrodynamics configuration the anharmonic quantum device is strongly coupled with an electromagnetic transmission line resonator, which allows the quantum non-demolition measurement of the state of the transmon by probing resonator's resonance frequency with microwaves.

%Superconducting solid-state quantum devices are very promising building blocks for quantum processors due to their scalability and long coherence times MAKE SURE THIS IS THE REASON. Our system consists of a transmon strongly coupled with an electromagnetic transmission line resonator. This architecture allows the quantum non-demolition measurement of the state of the transmon by probing the state dependant shift in the resonator frequency. On the other hand, the state of the system can be changed by applying a microwave drive on the gate of the transmon.

%Precise transfer of population from one quantum state to another is an integral part of quantum information processing.
%In a system with anharmonic energy spectrum coherent transfer of population can be achieved by

 In the transmon, it is not possible to transfer directly the population between the ground state $\ket{0}$ and the second excited state $\ket{2}$ due to the fact that the electric dipole momentum between these two states is vanishingly small. Thus this transfer must be done by first populating the first excited state $\ket{1}$ using a $\pi$ pulse applied to the $0-1$ transition, then transferring this population to the second excited state $\ket{2}$ using another $\pi$ pulse applied to the $1-2$ transition.
 This is called an intuitive sequence, and requires a precise control of the pulse sequence, as well as of the frequencies of the microwaves. In contrast, when the transfer is realized using STIRAP, there is no population in the intermediate state during the process. Because of this, the decay associated with the intermediate state does not matter. The population transfer in STIRAP is based on adiabatically varying control microwave fields, applied in a counter-intuitive order, {\it i.e.} with the first pulse driving the 12-transition, and the second pulse, having a suitable overlap with the first one, driving the 01-transition. The adiabatic nature of the population transfer makes the protocol very robust to the timing and the amplitudes of the pulses, as well as to the frequencies of the microwaves. Ideally, to ensure adiabaticity one should use long pulses. However, we find that this is not a strong restrictions, and the protocol can be run relatively fast using shorter pulses.

We consider a transmon driven by two microwave pulses of amplitudes $\Omega_{01}(t)$ and $\Omega_{12}(t)$ with frequencies
$\omega_{01}^{(\Omega )}(t)$ and $\omega_{12}^{(\Omega )}(t)$, possibly slightly detuned from the corresponding qubit transition frequencies $\omega_{01}$ and
$\omega_{12}$. In the dispersive regime and in the rotating wave approximation with respect to the two driving tones (see Supplementary Information section I), the effective three-level Hamiltonian of the driven transmon reads \cite{AT_us,nonabelianwallraff}
% The transitions between these three states are performed by two drive signals, both having different tone. In the rotating wave approximation (RWA) the Hamiltonian of the system can be written
\begin{equation}
\label{eq:awg_hamiltonian}
\op{H}(t) = \frac{\hbar}{2}
\begin{bmatrix}
0 & \Omega_{01}(t) & 0 \\
\Omega_{01}(t) & 2\delta_{01} & \Omega_{12}(t) \\
0 & \Omega_{12}(t) & 2(\delta_{01} + \delta_{12})
\end{bmatrix},
\end{equation}
where the detunings $\delta_{01}=\tilde{\omega}_{10} - \omega_{01}^{(\Omega )}$ and
$\delta_{12} = \tilde{\omega}_{21} - \omega_{12}^{(\Omega )}$
are defined with respect to the Lamb-shift renormalized transition frequencies of the transmon $\tilde{\omega}_{01} = \omega_{01} + \chi_{01}$  and $\tilde{\omega}_{12} = \omega_{12} + \chi_{12} - \chi_{01} $, and for simplicity $\Omega_{01}$ and $\Omega_{12}$ are taken real. The rates $\chi_{01}$ and $\chi_{12}$ are the ac-Stark shifts of the corresponding transitions.
%where $\Omega_{0,1}$ and $\Omega_{12}$ are the Rabi frequencies of the driven  $\ket{0}\rightarrow\ket{1}$ and $\ket{1}\rightarrow\ket{2}$ transitions, respectively. The parameters $\delta{0,1}$ and $\delta_{1,2}$ describe the detunings of the drive signals from the corresponding transitions.
If the two-photon resonant condition $\delta_{01} + \delta_{12}=0$ is satisfied, the Hamiltonian Eq. (\ref{eq:awg_hamiltonian}) has a zero-eigenvalue eigenstate called dark state,
$\ket{\rm D} = \cos{\Theta}\ket{0} - \sin{\Theta}\ket{2}$ (see Methods), with $\Theta$ defined by $\tan{\Theta} = \Omega_{01}(t)/\Omega_{12}(t)$. Due to the adiabatic theorem, by slowly tuning the drive amplitudes $\Omega_{01}(t)$ and $\Omega_{12}(t)$, such that $\tan{\Theta}$ goes from zero to infinity, the population transfer from the ground state to the second excited state can be realized.

In the experiment we work with Gaussian pulses, parametrized as
\begin{eqnarray}
\label{eq:gaussian_envelopes}
%\begin{aligned}
%\Omega_{01}(t) = \Omega_{01}\exp\left[-\frac{(t - t_{s}/2)^2}{2\sigma^2}\right], \\
%\Omega_{12}(t) = \Omega_{12}\exp\left[-\frac{(t + t_{s}/2)^2}{2\sigma^2}\right], \\
\Omega_{01}(t) &=& \Omega_{01}\exp\left[-\frac{t^2}{2\sigma^2}\right], \\
\Omega_{12}(t) &=& \Omega_{12}\exp\left[-\frac{(t - t_{s}/2)^2}{2\sigma^2}\right],
%\end{aligned}
\end{eqnarray}
where $t_{s}$ is the time separation between the maxima of the pulses, $\sigma$ is the width of the pulse, and $\Omega_{01}$, $\Omega_{12}$ are the amplitudes. We choose equal-amplitude pulses, $\Omega_{01} \approx \Omega_{12}\approx \Omega$.
%Note that, differently from holonomic gates, in our experiment the ratio $\Omega_{12}(t)/\Omega_{01}(t)$ varies in time.
The general condition for the adiabatic following is given by \cite{reviewSTIRAP,another_reviewSTIRAP}
$\abs{\bra{\pm}(d/dt)\ket{\rm D}}\ll \abs{\omega_\pm - \omega_{\rm D}}$.
%\left|\frac{\bra{\psi_m}\dot{\op{H}}\ket{\psi_n}}{E_n - E_m}\right| \ll 1, % Might not be correct. In other articles the condition is without H
By using \eqref{eq:gaussian_envelopes} and performing a time integration (Supplementary Information section II), we obtain the global condition of adiabaticity for Gaussian pulses,
\begin{equation}
\label{eq:global_adiabatic_condition}
\frac{4}{\sqrt{\pi}}\sigma\Omega \gg 1.
\end{equation}

This global adiabatic condition shows that the transfer efficiency can be improved by making the pulses longer, which is in turn limited by the decoherence of the states $\ket{0}$ and $\ket{2}$. In principle, if a high transfer efficiency is desired, the shape of the pulses could be further optimized \cite{optimal_stirap_pulses,composite_stirap}.

%In a real experiment the adiabatic condition is never fully satisfied, which leads to some population in the first excited state.

\section*{Experimental results}

From the spectroscopy data as well as from the change of the Rabi frequency as a function of the detuning, we determine the Lamb-shift renormalized transition frequencies of the transmon $\tilde{\omega}_{01}/2\pi = $ 5.27 GHz and $\tilde{\omega}_{12}/2\pi = $ 4.82 GHz. The dissipation rates extracted from independent measurements were $\Gamma_{10} =$ 2.4 MHz, $\Gamma_{21} = 5.2$ MHz, %$\Gamma_{21}/2\pi =$ 5.4 MHz.
and $\Gamma_{21}^{\varphi} \approx \Gamma_{10}^{\varphi} = $ 0.4 MHz, while the Gaussian pulses were calibrated to $\Omega_{01}/2\pi = $ 43.4 MHz, and $\Omega_{12}/2\pi = $ 38.2 MHz, with pulse width $\sigma = 45 $ ns for both pulses (see Methods). This choice satisfies the adiabatic condition in Eq. \eqref{eq:global_adiabatic_condition} by a factor of 4.

\begin{figure}[tbp]
\centering
\includegraphics[width = 1.0\columnwidth]{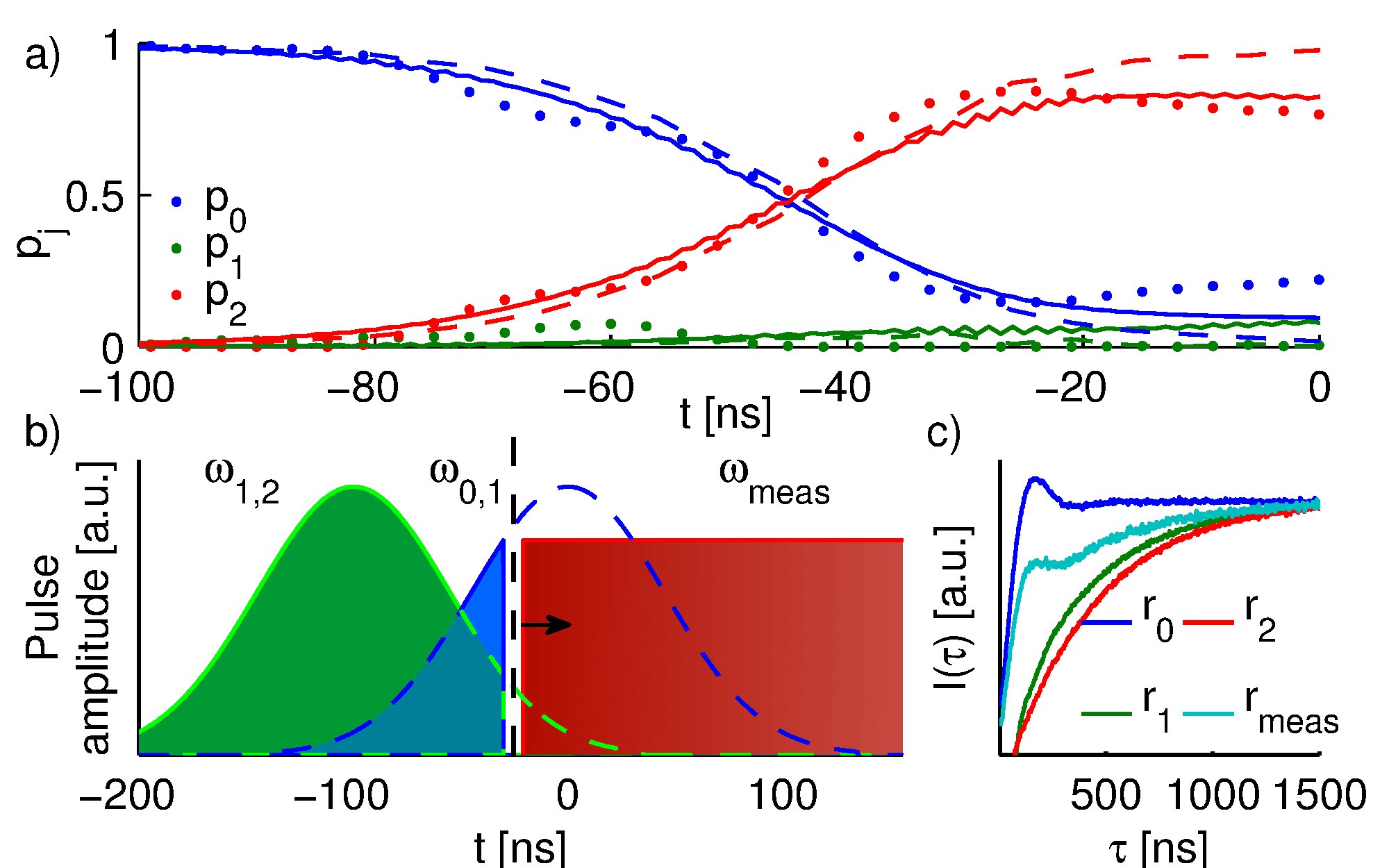}
\caption{a) The state of the quantum system at different times during the adiabatic population transfer at fixed pulse separation. The dots represent experimental results, the dashed line is a simulation of the three-level dissipationless case, and the continuos lines show the simulation with dissipation included. During the time evolution the intermediate state $\ket{1}$ remains almost unpopulated. b) Schematic of the pulse sequence during the STIRAP. %The time resolved plot in a) consists of many fractional STIRAP experiments with the  mixing angle increasing at each time step. After the STIRAP pulses the state of the system is measured by probing the cavity with a measurement pulse.
c) The population of the system at every time $t$ is calculated from the cavity response $r(\tau ) = \{ I(\tau ), Q(\tau )\}$ (only $I(\tau )$ shown here), and the measurement data (cyan) is compared to the calibration data for the known states $\ket{0}$ (blue),$\ket{1}$ (green) and $\ket{2}$ (red).}
\label{fig:stirap_measurement}
\end{figure}

Fig. \ref{fig:stirap_measurement}a) demonstrates the STIRAP protocol for our system, with the population of the  state $|2\rangle$ reaching a maximum value during the pulse overlap and then slowly decreasing at large timescales due to the intrinsic decay of the second excited state. The pulse sequence is presented in Fig. \ref{fig:stirap_measurement}b). The results from a quantum tomography measurement for three levels is shown schematically in Fig. \ref{fig:stirap_measurement}c). The traces are obtained by a homodyne measurement of the cavity response $r_j(\tau) = \{\avg{I_j(\tau)},\avg{Q_j(\tau)}\}$, $j =0,1,2$ for the reference states $\ket{0}$, $\ket{1}$, and $\ket{2}$, which are prepared by using fast microwave pulses applied to the 0-1 and 1-2 transition (Supplementary Information section III). Here $I_j$ and $Q_j$ are the in-phase and the quadrature fields. For an arbitrary state $\rho (t)$ at time $t$ during the STIRAP protocol the cavity response is a linear combination \cite{three_level_tomography_bianchetti}
$r_{\rm meas}(t,\tau) = \sum_{j=0,1,2} p_{j}(t) r_{j} (\tau )$, where $p_{j}(t) = {\rm Tr} [\rho (t)\ket{j}\bra{j}]$ is the population of state $|j\rangle$. We extract the populations $p_{j}(t)$ using the
Levenberg-Marquardt algorithm. We note also that the tomography method presented here includes interrupting the STIRAP protocol before the full population is transferred, a protocol called fractional STIRAP \cite{fractional_stirap}.

From the data we can see that the population in the second excited state reaches about 83\% and the
intermediate state gets slowly populated due to the decay $\ket{2} \rightarrow \ket{1}$. Thus, the main sources of nonideality are the decay and the dephasing of the upper levels, as well as imperfections in the timing of the short pulses used for calibration. A surprising finding is that the STIRAP transfer occurs in fact quite fast, predominantly in the overlap region: the optimal value is obtained not too far from the beginning of the $\omega_{01}$ pulse. This means that the duration of the STIRAP process can be shortened by interrupting it earlier (with correspondingly shaped pulses) without
too big penalties on the fidelity of state transfer, thereby reducing the overall gate length.

\begin{figure}[tbp]
%%\captionsetup[subfigure]{labelformat=empty}
%%\subfloat[]
\includegraphics[width=1.0\columnwidth]{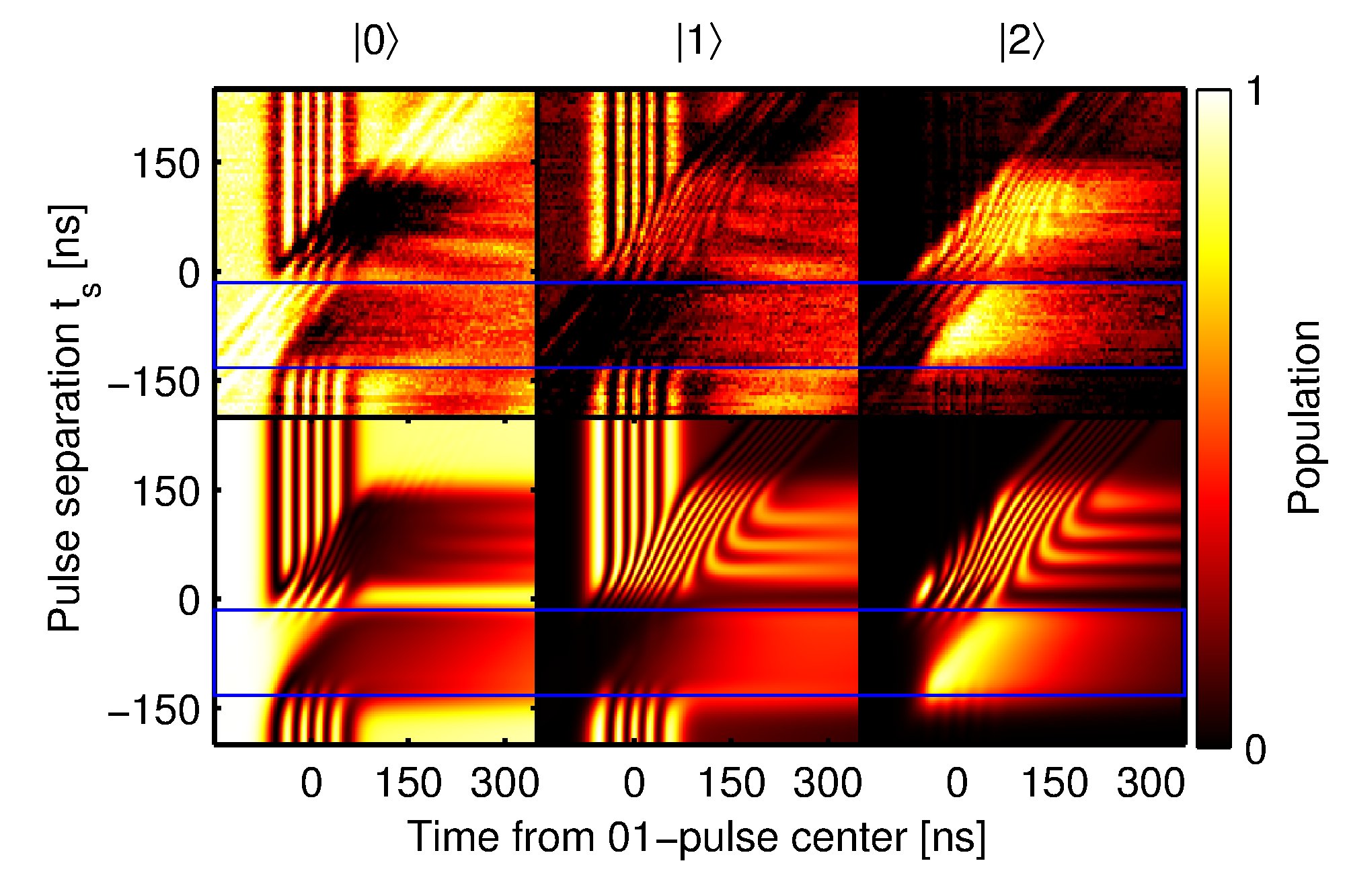}%\label{fig:pulse_overlap}}%\hfill
%%\subfloat[]
%\includegraphics[width=1.0\columnwidth]{./Figures/stirap_single_sweep_with_theory_combined}
%%\label{fig:stirap_single_sweep_with_theory}}
%%\subfloat[]{\includegraphics[width=1.0\textwidth]{stirap_single_sweep_with_theory}\label{fig:stirap_single_sweep_with_theory}}%
\caption{Time-resolved two-pulse sequence as a function of the $\omega_{01}$ and $\omega_{12}$ pulse separation, where 0 means that the pulses are completely overlapping.
%The standard deviation of the used Gaussian pulses is $\sigma = 45$ ns.
The upper figures show the experimental result for the time evolution of states $\ket{0}$,$\ket{1}$ and $\ket{2}$, whereas the lower row shows the corresponding simulation results, calculated using Lindblad's master equation for a multilevel Jaynes-Cummings Hamiltonian (see Methods). The STIRAP sequence (blue rectangle) occurs at negative values of the pulse separation, while positive values of the pulse separation correspond to the intuitive sequence.}
%b)Time evolution of the populations for the counterintuitive sequence (STIRAP) at optimal pulse separation of $-90$ ns, corresponding to the magenta arrows in a).
%c)Time evolution of the populations for an intuitive pulse sequence at a pulse separation of $90$ ns, indicated with green arrows in a).
%d)Populations on different states with a fixed delay of $-100$ ns of the 12 pulse with respect to the measurement pulse as a function of the 12 to 01 pulse separation, corresponding to a slice along the blue arrows in a).}
\label{fig:pulse_overlap}
\end{figure}

Next, we present a complete experimental investigation of this effect, together with numerical simulations. By varying the time separation between the two pulses, one can study the counterintuitive as well as the intuitive sequence. In Fig. \ref{fig:pulse_overlap} we present the occupation probability of each of the states $|0\rangle$, $|1\rangle$, and $|2\rangle$ at different measurement times (horizontal axis), measured from the peak of the 01 pulse.  The pulse separation (vertical axis) is the time interval between the peak of the 12 pulse and the peak of the 01 pulse. Thus, zero pulse separation means complete pulse overlap, negative values of the pulse separation refer to the counter-intuitive pulse sequence ($\omega_{12}$ is applied before $\omega_{01}$ pulse) whereas positive values indicate the pulses are send in the intuitive order. The lower panels of Fig. \ref{fig:pulse_overlap} show the corresponding simulation result using Lindblad's master equation (see Eq. (\ref{lindblad}) in Methods).  From the experimental data we find that the optimal value for the pulse separation is $\approx -90$ ns, in close agreement with the result given by \cite{optimal_stirap_pulses} and our own simulations.
Importantly, one notices that for STIRAP the transfer efficiency is high and a relatively slowly varying function of the pulse overlap in a plateau extending from about -120 ns to about -80 ns, which demonstrates that accurate optimization of the pulse overlap is not required.
%Fig. \ref{fig:pulse_overlap}b) shows a slice along the magenta arrows in Fig.\ref{fig:pulse_overlap}a) %at a constant pulse separation of -90 ns.
%From the data we extract the fidelity of the population
%transfer to be 99 \%. However, according to the simulation, maximal achievable fidelity is only 83 \% with the chosen parameters and the energy relaxation rates. The
%discrepancy shows the limitations of the applied measurement scheme. Unavoidably, the calibration measurements are not perfect due to their incorrect timing, energy relaxation and frequency fluctuations. As a result,
%the acquired fidelity here is not an absolute measure, but
%it rather compares STIRAP to the method of coherently
%exciting the transmon used to perform the calibration.
%The achieved high value of fidelity shows that the adia-
%batic method works atleast equally well when compared
%to coherently exciting the transmon. The comparison to the analytical result given by Eq. \eqref{eq:awg_states}, and the simulation of the full system Hamiltonian in Eq. \eqref{eq:jc_multilevel} shows that the non-ideal fidelity is mostly due to the decoherence of the state $\ket{2}$.
In principle, population transfer can also be realized using the intuitive-pulse sequence, but in this case the method is sensitive to the timing of the pulses, as can be seen from the oscillations present at positive pulse separation in Fig. \ref{fig:pulse_overlap}. Clearly the stability plateaus of the STIRAP do not form in the case of the intuitive sequence.
%Fig. \ref{fig:pulse_overlap}c) demonstrates the adiabatic population transfer using the intuitive pulse sequence.
Moreover, in the presence of decoherence it is harder to achieve transfer fidelities as high as with counter-intuitive pulse sequence, because the pulse sequence cannot be reliably interrupted until it is completely finished.
%Finally, Fig. \ref{fig:pulse_overlap}d) presents the population transfer with the 12 pulse fixed and followed by the measurement pulse after 100 ns, and with the position of the 01 pulse variable. The horizontal scale shows the separation between the 12 and 01 pulses; this corresponds to a diagonal slice along the blue arrows, as indicated in Fig. \ref{fig:pulse_overlap}a). This plot demonstrates the robustness of STIRAP (negative pulse separation) compared to the intuitive sequence (positive pulse separation): for STIRAP the transfer efficiency is high and a relatively slowly varying function of the pulse overlap in a plateau extending from about -120 ns to about -80 ns, which demonstrates that accurate optimization of the pulse overlap is not required. This is not the case at positive values of the separation, where oscillations appear.
%This is not the case at positive values of the separation, where oscillations appear.
This robustness against errors in the timing of the pulses demonstrated here
is one of the most significant benefits of STIRAP over non-adiabatic methods of population transfer.

% demonstrates this by showing the time evolution of the system states during STIRAP for various different values for the pulse overlap. In the experiment, the durations of the drive pulses were $\sigma =$ 50 ns. The optimal pulse separation is 70 ns, which is in close agreement with the result given in \cite{optimal_pulse_overlap}.

%shows the plateau of high efficiency, which is a signature of the STIRAP process. The width of the plateau makes STIRAP very robust against minor fluctuations in timing of the pulses.

\begin{figure}[tbp]
\centering
\includegraphics[width = 1.0\columnwidth]{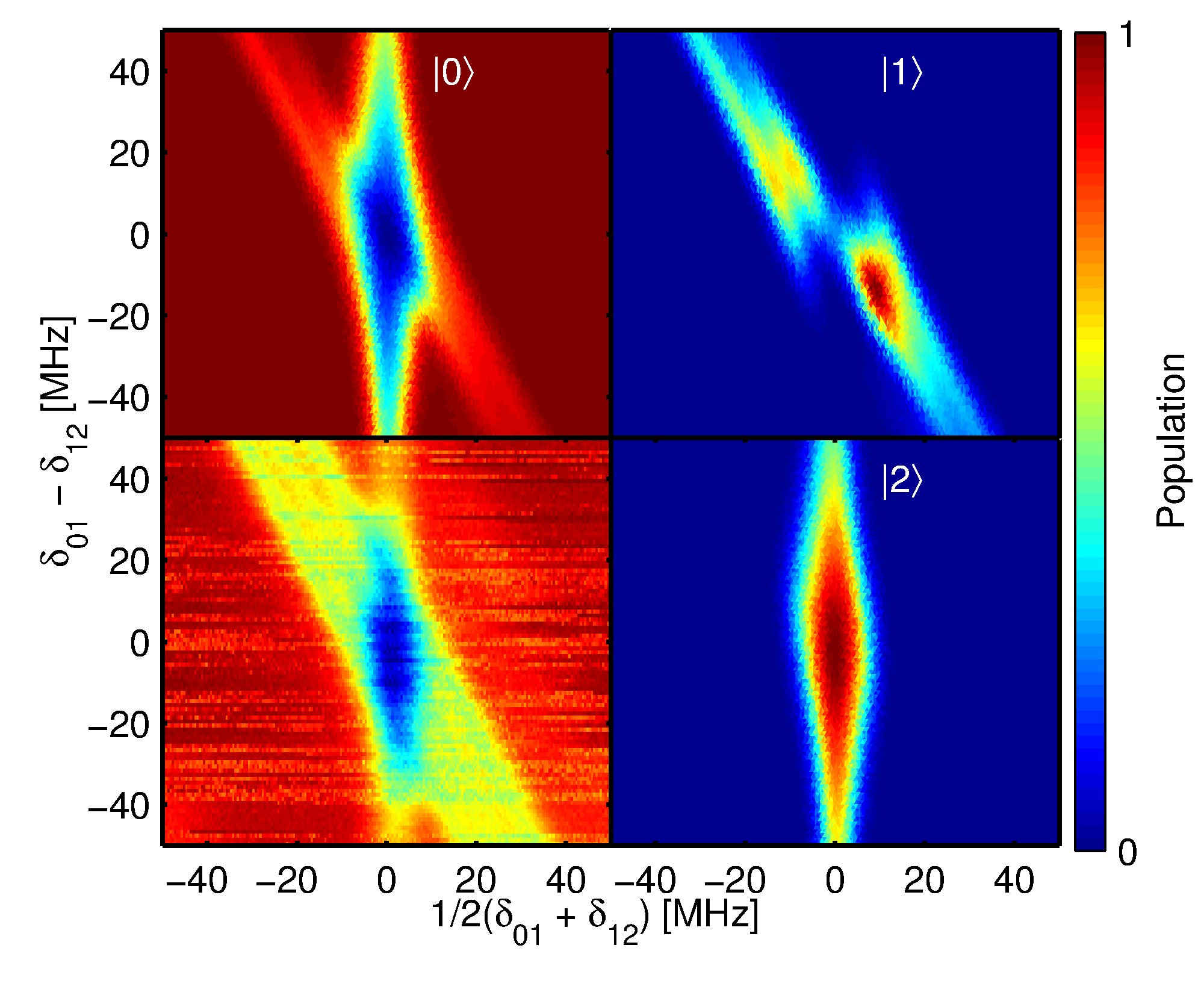}
\caption{STIRAP as a function of the detunings of the drive pulses. The offset from zero of the sum of the detunings $\delta_{01} +  \delta_{12}$ corresponds to the violation of the two photon resonance condition, to which the transfer is mostly sensitive. Nonzero values of the difference of the detunings $\delta_{01} -  \delta_{12}$ correspond to the violation of the single photon detuning, against which the system is robust. The lower left corner is the measured ground state population, while the corresponding simulation results are shown in the other plots.}
\label{fig:diamond}
\end{figure}

STIRAP is also very robust with respect to small errors in the frequencies of the drive signals. The effect is most notable along the axis where the two photon resonance condition is not violated, as demonstrated in Fig. \ref{fig:diamond}. This makes the method resilient for example to frequency shifts that occur in qubits due to charge and flux fluctuations.  Finally, STIRAP is not very sensitive to the widths of the pulses: we have tested $\sigma = 40,60,70$ ns and obtained similarly large values for the transfer efficiency. We find that when decreasing the pulse width below these values the adiabaticity condition is no longer fulfilled and oscillation start to appear, while if it is too large the effect of decay becomes significant.

%In the experiment we only measured the population of the $\ket{0}$ state, but comparison to the simulation using Eq. \eqref{eq:jc_multilevel} allows us to identify the parts from the measurement data which correspond to the $\ket{2}$ state.

%This is demonstrated in Fig. \ref{fig:diamond}.

%This is especially true for one photon detuning,

% which is shown in Fig. \ref{fig:diamond}. However, when the two photon resonance condition is violated, the efficiency of population transfer quickly diminishes.

\begin{figure}[tbp]
\centering
\includegraphics[width = 1.0\columnwidth]{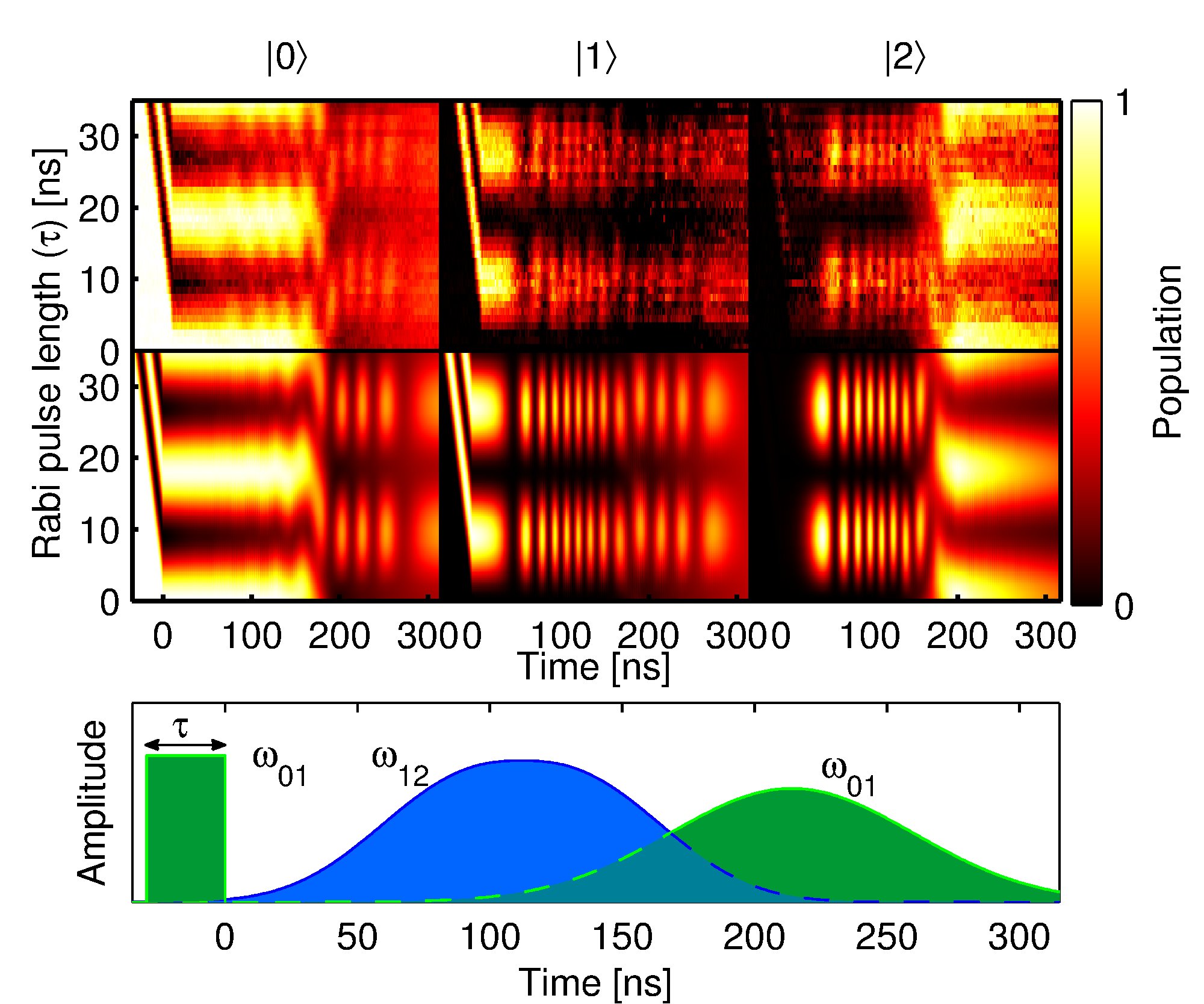}
\caption{a) The STIRAP efficiency depends on the initial population of the ground state. The Rabi pulse before the STIRAP pulse sequence transfers some of the population to the intermediate state $\ket{1}$, which reduces the population of level $|2\rangle$ at the end of the pulse sequence. The upper row shows the measurement results for the three states. The corresponding simulation results are shown on the lower row. b) The pulse sequence used to generate the measurement results.}
\label{fig:state_prep}
\end{figure}

During the operation of a quantum processor, it might be efficient to use both fast pulses and adiabatic pulses. Thus it is interesting to look at hybrid pulses, where for example a fast pulse is applied to the transition 01 before a STIRAP sequence, see Fig. \ref{fig:state_prep}. This corresponds to starting the adiabatic population transfer protocol from a superposition between the ground state and the first excited state. During the evolution of the system, this results in a superposition between the dark state and the intermediate state, as shown in Fig. \ref{fig:state_prep}. Towards the end of the sequence, the population on level 2 is stabilized (to a reduced values) but the populations of states 0 and 1 oscillate due to the Rabi coupling provided by the last $\omega_{01}$ pulse.

%\begin{figure}[tbp]
%\centering
%\includegraphics[width=1.0\columnwidth]{./Figures/three_gaussians}
%excited state then bring it back to the ground state. We present
%a) the time evolution of the state, b) the applied pulse sequence, and  c) the evolution of the system when the third Gaussian $\omega_{01}$-pulse is separated by $\tau_{\rm 3G} = 120$ ns from the middle $\omega_{12}$ pulse.}
%\label{fig:three_gaussians}
%\end{figure}

\begin{figure}[tbp]
\centering
\includegraphics[width=1.0\columnwidth]{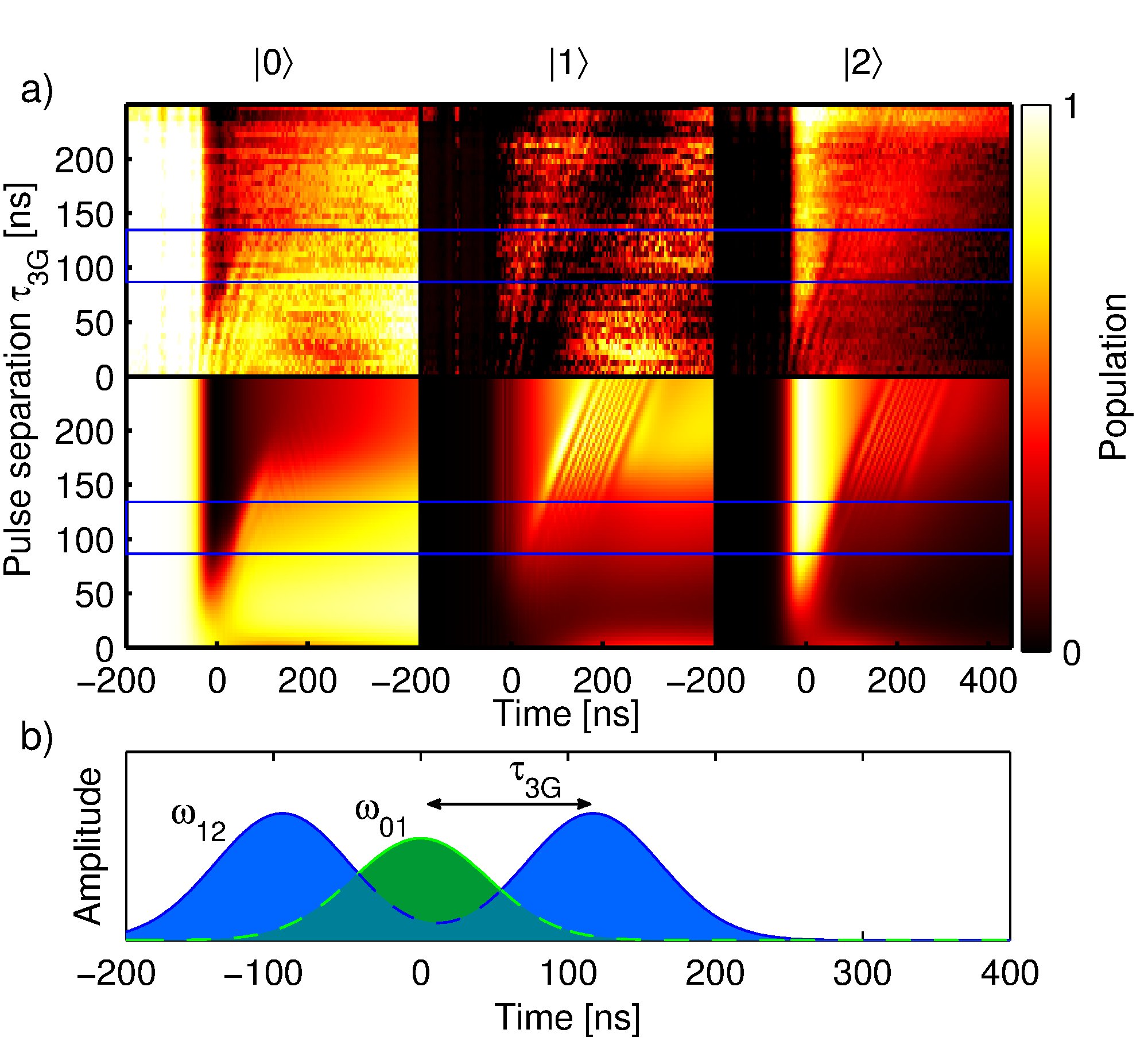}
\caption{STIRAP reversal: three adiabatic pulses are used to first excite the system to the second excited state and then bring it back to the ground state.
a) The time evolution of the state (with the blue rectangle indicating the region where the process is the most efficient). The upper row is experiment, while the lower row is simulation.  b) Schematic of the applied pulse sequence.}
\label{fig:three_gaussians}
\end{figure}

In Fig. \ref{fig:three_gaussians}a) we show how the STIRAP can be reversed by applying one additional Gaussian pulse at the $\omega_{12}$ transition. The reversal of the STIRAP is essential in creating adiabatic gates, for example to realizing a single-qubit phase gate \cite{geometricphasegate}.
The idea is to first excite the system to state $\ket{2}$ using STIRAP, and then directly reapply STIRAP to bring the system back to the ground state. This demonstrates that STIRAP can also be used to transfer population from state $\ket{2}$ back to ground state.
 %However, the procedure of preparing the state and then reapplying STIRAP takes relatively long when compared to the coherence time of our system reducing the fidelity of the operation quite significantly.

\begin{figure}[tbp]
\centering
\includegraphics[width = 1.0\columnwidth]{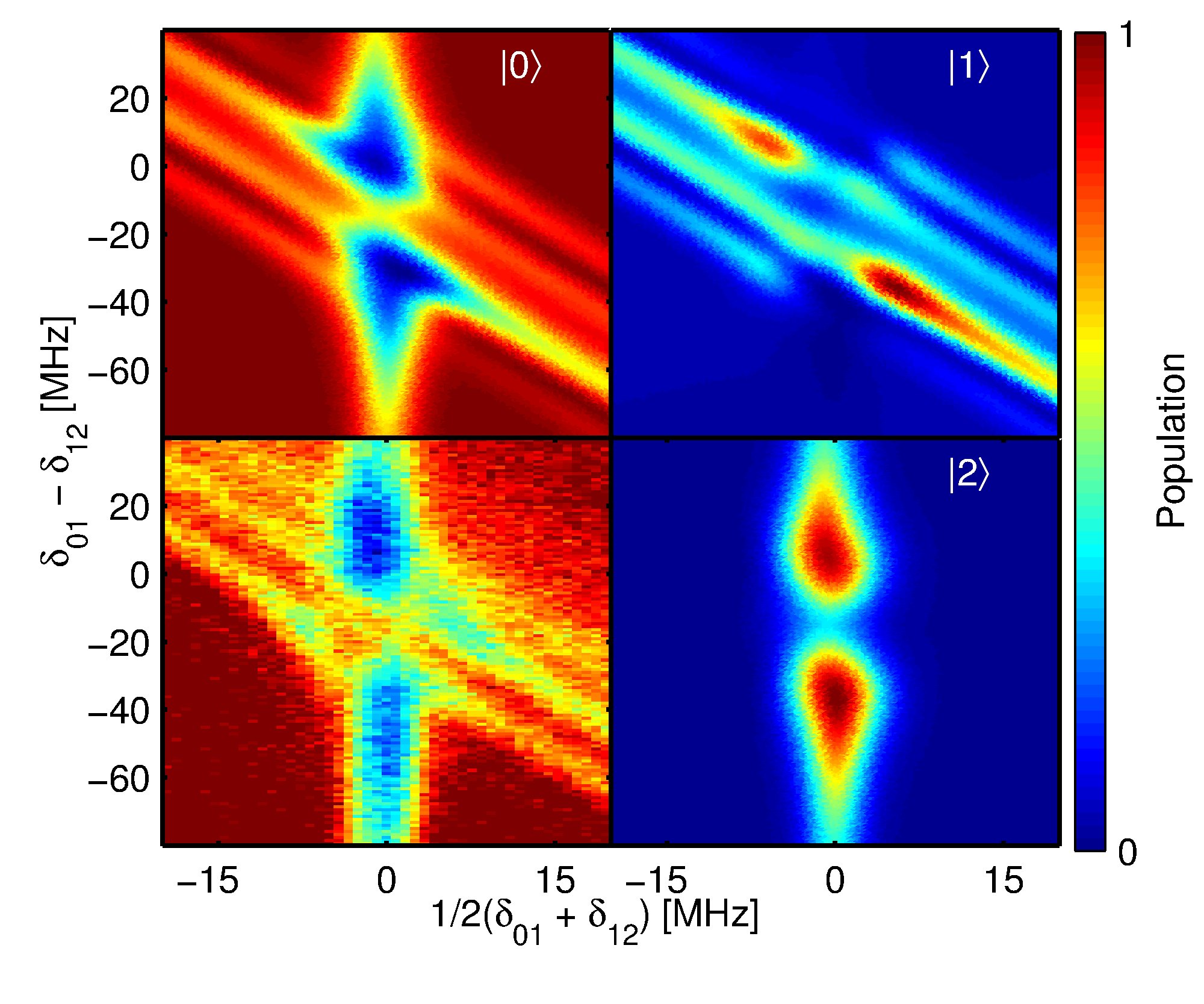}
\caption{The splitting of the first excited state to the states $\ket{1_a}$ and $\ket{1_b}$ suppresses the transfer efficiency when the one photon detuning is between the split levels. The energy separation between the levels is 15 MHz.}
\label{fig:split_diamond}
\end{figure}

Finally, an interesting situation is that of a degenerate or quasi-degenerate intermediate state. To realize this degeneracy we did several thermal cycles of the sample, until we created a defect in the insulating layer of the junctions, observed by the splitting of the spectral line into states $\ket{1_a}$ and $\ket{1_b}$. The formation of this type of defect has been observed recently in flux qubits as well \cite{lupascu}. In this case an interesting two-blob structure can be seen in the detuning plot, as shown in Fig. \ref{fig:split_diamond}, where each blob corresponds to the resonance conditions of either $\ket{1_a}$ or $\ket{1_b}$.

% The dynamics of split energy levels cannot be explained by the Hamiltonian in Eq. \eqref{eq:awg_hamiltonian} due to the breakdown of the rotating wave approximation, but must be numerically solved from the bare Hamiltonian.

\section*{Discussion and conclusions}
We have demonstrated the adiabatic transfer of population from the ground state $\ket{0}$ to the second excited state $\ket{2}$ in a three level transmon circuit.
This benchmarks the STIRAP protocol as a valid procedure for quantum information processing applications in circuit QED. As an added bonus, we note that while in the previous continuous-wave experiments on various types of qubits \cite{AT_us,AT_wallraff,coherent_population_trapping,EIT_abdumalikov} it was not possible to asses unambiguosly the effect of quantum interference from the spectrocopic data \cite{sanders}, our experiment establishes it conclusively. We also studied the effect of having a finite initial state population in the dark state, we demonstrated the adiabatic reversal of the STIRAP protocol, and we realized the transfer of population also via a quasi-degenerate intermediate state. The main limiting factor is decoherence, which however for superconducting qubits is improving fast with new designs and fabrication techniques. We find that the process is very resilient to errors in the timing of the pulses and the frequencies of the microwave drives.

\section*{Methods}

The effective Hamiltonian of the system in a doubly rotating frame and under the rotating-wave approximation takes the form  Eq. (\ref{eq:awg_hamiltonian}). If the two-photon detuning $\delta_{01} + \delta_{12}$ is zero, the eigenvalues of the above Hamiltonian are $\omega_+ = \delta_{01} + \sqrt{\delta_{01}^2 + \Omega_{01}^2 + \Omega_{12}^2}$, $\omega_- = \delta_{01} - \sqrt{\delta_{01}^2 + \Omega_{01}^2 + \Omega_{12}^2}$,
and $\omega_{\rm D} = 0$, with corresponding eigenvectors
%\begin{equation}
%\label{eq:awg_eigenvalues}
%\begin{aligned}
%\omega_+ &= \delta_{01} + \sqrt{\delta_{01}^2 + \Omega_{01}^2 + \Omega_{12}^2}, \\
%\omega_- &= \delta_{01} - \sqrt{\delta_{01}^2 + \Omega_{01}^2 + \Omega_{12}^2}, \\
%\omega_0 &= 0.
%\end{aligned}
%\end{equation}
%\begin{equation}
%\begin{aligned}
%\cos{\Theta} &= \frac{\Omega_{12}(t)}{\sqrt{\Omega_{01}(t)^2 + \Omega_{12}(t)^2}},\, \tan{\Theta} &= %\frac{\Omega_{01}(t)}{\Omega_{12}(t)} \\
%\sin{\Theta} &= \frac{\Omega_{01}(t)}{\sqrt{\Omega_{01}(t)^2 + \Omega_{12}(t)^2}}, \\
%\end{aligned}
%\end{equation}
\begin{equation}
\label{eq:adiabatic_states}
\begin{aligned}
\ket{+} &= \sin{\Phi}\ket{\rm B} + \cos{\Phi}\ket{1},\\
\ket{-} &= \cos{\Phi}\ket{\rm B} - \sin{\Phi}\ket{1},\\
\ket{\rm D} &= \cos{\Theta}\ket{0} - \sin{\Theta}\ket{2},
\end{aligned}
\end{equation}
where the zero-eigenvalue is called the dark state, and the state orthogonal to it $|{\rm B}\rangle = \sin{\Theta}\ket{0} + \cos{\Theta} \ket{2}$ is called the bright state.
The angle $\Theta$ is defined by $\tan{\Theta} = \Omega_{01}(t)/\Omega_{12}(t)$ and it parametrizes the rotation in the $\{|0\rangle, |2\rangle \}$ subspace (spanned also by the bright and dark states), while
\begin{equation}
  \Phi = \tan^{-1}\left[\frac{\sqrt{(\Omega_{01}^2 + \Omega_{12}^2)/2}}{\sqrt{(\Omega_{01}^2 + \Omega_{12}^2 + \delta_{01}^2)/2} + \delta_{01}/\sqrt{2}}\right].
\end{equation}
%$\Phi$ is an angle of a right triangle with vertices $\sqrt{(\Omega_{01}^2 + \Omega_{12}^2)/2}$ and $\sqrt{(\Omega_{01}^2 + \Omega_{12}^2 + \delta_{01}^2)/2} + \delta_{01}/\sqrt{2}$.
For zero-detuning $\delta_{01}=0$ we have $\Phi=\pi /4$.
The adiabatic tuning of the pulse amplitudes ensures that the state of the system follows the $\ket{\rm D}$ eigenstate instead of being nonadiabatically excited to either one of the $\ket{\pm}$ eigenstates, which contain contributions from the state $\ket{1}$.

The transition frequencies of the transmon $\tilde{\omega}_{01}/2\pi = $ 5.27 GHz, $\tilde{\omega}_{12}/2\pi = $ 4.82 GHz are obtained from the spectroscopy data as well as from the Chevron patterns of the detuned Rabi oscillations. The measurements of Rabi oscillations on the 1-2 transition are done by first applying a $\pi$ pulse to the 0-1 transition followed by the Rabi pulse on the 1-2 transition. To obtain the decay rate $\Gamma_{10}$ we used a $\pi$ pulse on the $0-1$ transition and measure the population of the level 1 as it decays. To find $\Gamma_{12}$ we use a $\pi$ pulse applied to the $0-1$ transition, then a $\pi$ pulse on the $1-2$ transition, and we fit the data with a three-level model which includes the decay $2\rightarrow 1$ and $1\rightarrow 0$. We obtain the energy relaxation rates $\Gamma_{10} =$ 2.4 MHz and $\Gamma_{21} = 5.2$ MHz. %$\Gamma_{21}/2\pi =$ 5.4 MHz.
From Ramsey interference experiments we get $\Gamma_{21}^{\varphi} \approx \Gamma_{10}^{\varphi} = $ 0.4 MHz.
The Gaussian pulses are created by a high-sample rate arbitrary waveform generator followed by mixing with a microwave pulse (see Supplementary information). They are calibrated using the Rabi data, resulting in $\Omega_{01}/2\pi = $ 43.4 MHz, and $\Omega_{12}/2\pi = $ 38.2 MHz, with standard deviation $\sigma = 45 $ ns for both pulses. The quantum tomography procedure follows Ref. \cite{three_level_tomography_bianchetti} (see Fig. \ref{fig:stirap_measurement}c) and the Supplementary information for details). In our experiment, the tomography pulse interrupts the STIRAP protocol before the full population is transferred, thus it could be extended as well to the investigation of arbitrary superpositions between the ground state and the second excited state in the so-called fractional STIRAP protocols \cite{fractional_stirap}. For the simulations we use the three-level Lindblad master equation
\begin{equation}
\frac{d}{dt}\op{\rho} = -\frac{i}{\hbar}\left[ \op{\rho}, \op{H}\right]+ {\cal L}_{\rm rel} [\rho ] + {\cal L}_{\rm deph} [\rho ], \label{lindblad}
\end{equation}
where the Lindblad superoperators are
\begin{eqnarray}
{\cal L}_{\rm rel} [\rho ] &=& \frac{\Gamma_{10}}{2}\left( 2 \sigma_{01} \rho \sigma_{01} - \sigma_{11} \rho
- \rho \sigma_{11} \right) \nonumber \\
& &+ \frac{\Gamma_{21}}{2}\left( 2 \sigma_{12} \rho \sigma_{21} - \sigma_{22} \rho
- \rho \sigma_{11} \right)
\end{eqnarray}
for the relaxation and
\begin{eqnarray}
{\cal L}_{\rm deph} [\rho ] &=& \frac{\Gamma_{10}^{\varphi}}{2}\left( 2 \sigma_{01} \rho \sigma_{01} - \sigma_{11} \rho
- \rho \sigma_{11} \right) \nonumber \\
& & + \frac{\Gamma_{21}^{\varphi}}{2}\left( 2 \sigma_{12} \rho \sigma_{21} - \sigma_{22} \rho
- \rho \sigma_{11} \right)
\end{eqnarray}
for the pure dephasing term and $\sigma_{ij} = \ket{i}\bra{j}$.
%
% ****** End of file apssamp.tex ******

\bibliography{ref}

%merlin.mbs apsrev4-1.bst 2010-07-25 4.21a (PWD, AO, DPC) hacked
%Control: key (0)
%Control: author (8) initials jnrlst
%Control: editor formatted (1) identically to author
%Control: production of article title (-1) disabled
%Control: page (0) single
%Control: year (1) truncated
%Control: production of eprint (0) enabled
\begin{thebibliography}{30}%
\makeatletter
\providecommand \@ifxundefined [1]{%
 \@ifx{#1\undefined}
}%
\providecommand \@ifnum [1]{%
 \ifnum #1\expandafter \@firstoftwo
 \else \expandafter \@secondoftwo
 \fi
}%
\providecommand \@ifx [1]{%
 \ifx #1\expandafter \@firstoftwo
 \else \expandafter \@secondoftwo
 \fi
}%
\providecommand \natexlab [1]{#1}%
\providecommand \enquote  [1]{``#1''}%
\providecommand \bibnamefont  [1]{#1}%
\providecommand \bibfnamefont [1]{#1}%
\providecommand \citenamefont [1]{#1}%
\providecommand \href@noop [0]{\@secondoftwo}%
\providecommand \href [0]{\begingroup \@sanitize@url \@href}%
\providecommand \@href[1]{\@@startlink{#1}\@@href}%
\providecommand \@@href[1]{\endgroup#1\@@endlink}%
\providecommand \@sanitize@url [0]{\catcode `\\12\catcode `\$12\catcode
  `\&12\catcode `\#12\catcode `\^12\catcode `\_12\catcode `\%12\relax}%
\providecommand \@@startlink[1]{}%
\providecommand \@@endlink[0]{}%
\providecommand \url  [0]{\begingroup\@sanitize@url \@url }%
\providecommand \@url [1]{\endgroup\@href {#1}{\urlprefix }}%
\providecommand \urlprefix  [0]{URL }%
\providecommand \Eprint [0]{\href }%
\providecommand \doibase [0]{http://dx.doi.org/}%
\providecommand \selectlanguage [0]{\@gobble}%
\providecommand \bibinfo  [0]{\@secondoftwo}%
\providecommand \bibfield  [0]{\@secondoftwo}%
\providecommand \translation [1]{[#1]}%
\providecommand \BibitemOpen [0]{}%
\providecommand \bibitemStop [0]{}%
\providecommand \bibitemNoStop [0]{.\EOS\space}%
\providecommand \EOS [0]{\spacefactor3000\relax}%
\providecommand \BibitemShut  [1]{\csname bibitem#1\endcsname}%
\let\auto@bib@innerbib\@empty
%</preamble>
\bibitem [{\citenamefont {Lanyon}\ \emph {et~al.}(2009)\citenamefont {Lanyon},
  \citenamefont {Barbieri}, \citenamefont {Almeida}, \citenamefont {Jennewein},
  \citenamefont {Ralph}, \citenamefont {Resch}, \citenamefont {Pryde},
  \citenamefont {O'Brien}, \citenamefont {Gilchrist},\ and\ \citenamefont
  {White}}]{Lanyon2009}%
  \BibitemOpen
  \bibfield  {author} {\bibinfo {author} {\bibfnamefont {B.~P.}\ \bibnamefont
  {Lanyon}}, \bibinfo {author} {\bibfnamefont {M.}~\bibnamefont {Barbieri}},
  \bibinfo {author} {\bibfnamefont {M.~P.}\ \bibnamefont {Almeida}}, \bibinfo
  {author} {\bibfnamefont {T.}~\bibnamefont {Jennewein}}, \bibinfo {author}
  {\bibfnamefont {T.~C.}\ \bibnamefont {Ralph}}, \bibinfo {author}
  {\bibfnamefont {K.~J.}\ \bibnamefont {Resch}}, \bibinfo {author}
  {\bibfnamefont {G.~J.}\ \bibnamefont {Pryde}}, \bibinfo {author}
  {\bibfnamefont {J.~L.}\ \bibnamefont {O'Brien}}, \bibinfo {author}
  {\bibfnamefont {A.}~\bibnamefont {Gilchrist}}, \ and\ \bibinfo {author}
  {\bibfnamefont {A.~G.}\ \bibnamefont {White}},\ }\href {\doibase
  10.1038/nphys1150} {\bibfield  {journal} {\bibinfo  {journal} {Nat Phys}\
  }\textbf {\bibinfo {volume} {5}},\ \bibinfo {pages} {134} (\bibinfo {year}
  {2009})}\BibitemShut {NoStop}%
\bibitem [{\citenamefont {Sillanpää}\ \emph {et~al.}(2009)\citenamefont
  {Sillanpää}, \citenamefont {Li}, \citenamefont {Cicak}, \citenamefont
  {Altomare}, \citenamefont {Park}, \citenamefont {Simmonds}, \citenamefont
  {Paraoanu},\ and\ \citenamefont {Hakonen}}]{AT_us}%
  \BibitemOpen
  \bibfield  {author} {\bibinfo {author} {\bibfnamefont {M.~A.}\ \bibnamefont
  {Sillanpää}}, \bibinfo {author} {\bibfnamefont {J.}~\bibnamefont {Li}},
  \bibinfo {author} {\bibfnamefont {K.}~\bibnamefont {Cicak}}, \bibinfo
  {author} {\bibfnamefont {F.}~\bibnamefont {Altomare}}, \bibinfo {author}
  {\bibfnamefont {J.~I.}\ \bibnamefont {Park}}, \bibinfo {author}
  {\bibfnamefont {R.~W.}\ \bibnamefont {Simmonds}}, \bibinfo {author}
  {\bibfnamefont {G.~S.}\ \bibnamefont {Paraoanu}}, \ and\ \bibinfo {author}
  {\bibfnamefont {P.~J.}\ \bibnamefont {Hakonen}},\ }\href {\doibase
  10.1103/PhysRevLett.103.193601} {\bibfield  {journal} {\bibinfo  {journal}
  {Phys. Rev. Lett.}\ }\textbf {\bibinfo {volume} {103}},\ \bibinfo {pages}
  {193601} (\bibinfo {year} {2009})}\BibitemShut {NoStop}%
\bibitem [{\citenamefont {Baur}\ \emph {et~al.}(2009)\citenamefont {Baur},
  \citenamefont {Filipp}, \citenamefont {Bianchetti}, \citenamefont {Fink},
  \citenamefont {Göppl}, \citenamefont {Steffen}, \citenamefont {Leek},
  \citenamefont {Blais},\ and\ \citenamefont {Wallraff}}]{AT_wallraff}%
  \BibitemOpen
  \bibfield  {author} {\bibinfo {author} {\bibfnamefont {M.}~\bibnamefont
  {Baur}}, \bibinfo {author} {\bibfnamefont {S.}~\bibnamefont {Filipp}},
  \bibinfo {author} {\bibfnamefont {R.}~\bibnamefont {Bianchetti}}, \bibinfo
  {author} {\bibfnamefont {J.~M.}\ \bibnamefont {Fink}}, \bibinfo {author}
  {\bibfnamefont {M.}~\bibnamefont {Göppl}}, \bibinfo {author} {\bibfnamefont
  {L.}~\bibnamefont {Steffen}}, \bibinfo {author} {\bibfnamefont {P.~J.}\
  \bibnamefont {Leek}}, \bibinfo {author} {\bibfnamefont {A.}~\bibnamefont
  {Blais}}, \ and\ \bibinfo {author} {\bibfnamefont {A.}~\bibnamefont
  {Wallraff}},\ }\href {\doibase 10.1103/PhysRevLett.102.243602} {\bibfield
  {journal} {\bibinfo  {journal} {Phys. Rev. Lett.}\ }\textbf {\bibinfo
  {volume} {102}},\ \bibinfo {pages} {243602} (\bibinfo {year}
  {2009})}\BibitemShut {NoStop}%
\bibitem [{\citenamefont {Kelly}\ \emph {et~al.}(2010)\citenamefont {Kelly},
  \citenamefont {Dutton}, \citenamefont {Schlafer}, \citenamefont {Mookerji},
  \citenamefont {Ohki}, \citenamefont {Kline},\ and\ \citenamefont
  {Pappas}}]{coherent_population_trapping}%
  \BibitemOpen
  \bibfield  {author} {\bibinfo {author} {\bibfnamefont {W.~R.}\ \bibnamefont
  {Kelly}}, \bibinfo {author} {\bibfnamefont {Z.}~\bibnamefont {Dutton}},
  \bibinfo {author} {\bibfnamefont {J.}~\bibnamefont {Schlafer}}, \bibinfo
  {author} {\bibfnamefont {B.}~\bibnamefont {Mookerji}}, \bibinfo {author}
  {\bibfnamefont {T.~A.}\ \bibnamefont {Ohki}}, \bibinfo {author}
  {\bibfnamefont {J.~S.}\ \bibnamefont {Kline}}, \ and\ \bibinfo {author}
  {\bibfnamefont {D.~P.}\ \bibnamefont {Pappas}},\ }\href {\doibase
  10.1103/PhysRevLett.104.163601} {\bibfield  {journal} {\bibinfo  {journal}
  {Phys. Rev. Lett.}\ }\textbf {\bibinfo {volume} {104}},\ \bibinfo {pages}
  {163601} (\bibinfo {year} {2010})}\BibitemShut {NoStop}%
\bibitem [{\citenamefont {Abdumalikov}\ \emph {et~al.}(2010)\citenamefont
  {Abdumalikov}, \citenamefont {Astafiev}, \citenamefont {Zagoskin},
  \citenamefont {Pashkin}, \citenamefont {Nakamura},\ and\ \citenamefont
  {Tsai}}]{EIT_abdumalikov}%
  \BibitemOpen
  \bibfield  {author} {\bibinfo {author} {\bibfnamefont {A.~A.}\ \bibnamefont
  {Abdumalikov}}, \bibinfo {author} {\bibfnamefont {O.}~\bibnamefont
  {Astafiev}}, \bibinfo {author} {\bibfnamefont {A.~M.}\ \bibnamefont
  {Zagoskin}}, \bibinfo {author} {\bibfnamefont {Y.~A.}\ \bibnamefont
  {Pashkin}}, \bibinfo {author} {\bibfnamefont {Y.}~\bibnamefont {Nakamura}}, \
  and\ \bibinfo {author} {\bibfnamefont {J.~S.}\ \bibnamefont {Tsai}},\ }\href
  {\doibase 10.1103/PhysRevLett.104.193601} {\bibfield  {journal} {\bibinfo
  {journal} {Phys. Rev. Lett.}\ }\textbf {\bibinfo {volume} {104}},\ \bibinfo
  {pages} {193601} (\bibinfo {year} {2010})}\BibitemShut {NoStop}%
\bibitem [{\citenamefont {Leek}\ \emph {et~al.}(2007)\citenamefont {Leek},
  \citenamefont {Fink}, \citenamefont {Blais}, \citenamefont {Bianchetti},
  \citenamefont {G\"oppl}, \citenamefont {Gambetta}, \citenamefont {Schuster},
  \citenamefont {Frunzio}, \citenamefont {Schoelkopf},\ and\ \citenamefont
  {Wallraff}}]{berrywallraff}%
  \BibitemOpen
  \bibfield  {author} {\bibinfo {author} {\bibfnamefont {P.}~\bibnamefont
  {Leek}}, \bibinfo {author} {\bibfnamefont {J.~M.}\ \bibnamefont {Fink}},
  \bibinfo {author} {\bibfnamefont {A.}~\bibnamefont {Blais}}, \bibinfo
  {author} {\bibfnamefont {R.}~\bibnamefont {Bianchetti}}, \bibinfo {author}
  {\bibfnamefont {M.}~\bibnamefont {G\"oppl}}, \bibinfo {author} {\bibfnamefont
  {J.}~\bibnamefont {Gambetta}}, \bibinfo {author} {\bibfnamefont
  {D.}~\bibnamefont {Schuster}}, \bibinfo {author} {\bibfnamefont
  {L.}~\bibnamefont {Frunzio}}, \bibinfo {author} {\bibfnamefont
  {R.}~\bibnamefont {Schoelkopf}}, \ and\ \bibinfo {author} {\bibfnamefont
  {A.}~\bibnamefont {Wallraff}},\ }\href {\doibase 10.1126/science.1149858}
  {\bibfield  {journal} {\bibinfo  {journal} {Science}\ }\textbf {\bibinfo
  {volume} {318}},\ \bibinfo {pages} {1889} (\bibinfo {year}
  {2007})}\BibitemShut {NoStop}%
\bibitem [{\citenamefont {Abdumalikov}\ \emph {et~al.}(2013)\citenamefont
  {Abdumalikov}, \citenamefont {Fink}, \citenamefont {Juliusson}, \citenamefont
  {Pechal}, \citenamefont {Berger}, \citenamefont {Wallraff},\ and\
  \citenamefont {Filipp}}]{nonabelianwallraff}%
  \BibitemOpen
  \bibfield  {author} {\bibinfo {author} {\bibfnamefont {A.~A.}\ \bibnamefont
  {Abdumalikov}}, \bibinfo {author} {\bibfnamefont {J.~M.}\ \bibnamefont
  {Fink}}, \bibinfo {author} {\bibfnamefont {K.}~\bibnamefont {Juliusson}},
  \bibinfo {author} {\bibfnamefont {M.}~\bibnamefont {Pechal}}, \bibinfo
  {author} {\bibfnamefont {S.}~\bibnamefont {Berger}}, \bibinfo {author}
  {\bibfnamefont {A.}~\bibnamefont {Wallraff}}, \ and\ \bibinfo {author}
  {\bibfnamefont {S.}~\bibnamefont {Filipp}},\ }\href {\doibase
  10.1038/nature12010} {\bibfield  {journal} {\bibinfo  {journal} {Nature}\
  }\textbf {\bibinfo {volume} {496}},\ \bibinfo {pages} {482} (\bibinfo {year}
  {2013})}\BibitemShut {NoStop}%
\bibitem [{\citenamefont {Roushan}\ \emph {et~al.}(2014)\citenamefont
  {Roushan}, \citenamefont {Neill}, \citenamefont {Chen}, \citenamefont
  {Kolodrubetz}, \citenamefont {Quintana}, \citenamefont {Leung}, \citenamefont
  {Fang}, \citenamefont {Barends}, \citenamefont {Campbell}, \citenamefont
  {Chen}, \citenamefont {Chiaro}, \citenamefont {Dunsworth}, \citenamefont
  {Jeffrey}, \citenamefont {Kelly}, \citenamefont {Megrant}, \citenamefont
  {Mutus}, \citenamefont {O’Malley}, \citenamefont {Sank}, \citenamefont
  {Vainsencher}, \citenamefont {Wenner}, \citenamefont {White}, \citenamefont
  {Polkovnikov}, \citenamefont {Cleland},\ and\ \citenamefont
  {Martinis}}]{topologicaltransitions}%
  \BibitemOpen
  \bibfield  {author} {\bibinfo {author} {\bibfnamefont {P.}~\bibnamefont
  {Roushan}}, \bibinfo {author} {\bibfnamefont {C.}~\bibnamefont {Neill}},
  \bibinfo {author} {\bibfnamefont {Y.}~\bibnamefont {Chen}}, \bibinfo {author}
  {\bibfnamefont {M.}~\bibnamefont {Kolodrubetz}}, \bibinfo {author}
  {\bibfnamefont {C.}~\bibnamefont {Quintana}}, \bibinfo {author}
  {\bibfnamefont {N.}~\bibnamefont {Leung}}, \bibinfo {author} {\bibfnamefont
  {M.}~\bibnamefont {Fang}}, \bibinfo {author} {\bibfnamefont {R.}~\bibnamefont
  {Barends}}, \bibinfo {author} {\bibfnamefont {B.}~\bibnamefont {Campbell}},
  \bibinfo {author} {\bibfnamefont {Z.}~\bibnamefont {Chen}}, \bibinfo {author}
  {\bibfnamefont {B.}~\bibnamefont {Chiaro}}, \bibinfo {author} {\bibfnamefont
  {A.}~\bibnamefont {Dunsworth}}, \bibinfo {author} {\bibfnamefont
  {E.}~\bibnamefont {Jeffrey}}, \bibinfo {author} {\bibfnamefont
  {J.}~\bibnamefont {Kelly}}, \bibinfo {author} {\bibfnamefont
  {A.}~\bibnamefont {Megrant}}, \bibinfo {author} {\bibfnamefont
  {J.}~\bibnamefont {Mutus}}, \bibinfo {author} {\bibfnamefont {P.~J.~J.}\
  \bibnamefont {O’Malley}}, \bibinfo {author} {\bibfnamefont
  {D.}~\bibnamefont {Sank}}, \bibinfo {author} {\bibfnamefont {A.}~\bibnamefont
  {Vainsencher}}, \bibinfo {author} {\bibfnamefont {J.}~\bibnamefont {Wenner}},
  \bibinfo {author} {\bibfnamefont {T.}~\bibnamefont {White}}, \bibinfo
  {author} {\bibfnamefont {A.}~\bibnamefont {Polkovnikov}}, \bibinfo {author}
  {\bibfnamefont {A.~N.}\ \bibnamefont {Cleland}}, \ and\ \bibinfo {author}
  {\bibfnamefont {J.~M.}\ \bibnamefont {Martinis}},\ }\href {\doibase
  10.1038/nature13891} {\bibfield  {journal} {\bibinfo  {journal} {Nature}\
  }\textbf {\bibinfo {volume} {515}},\ \bibinfo {pages} {241} (\bibinfo {year}
  {2014})}\BibitemShut {NoStop}%
\bibitem [{\citenamefont {Das}\ and\ \citenamefont
  {Chakrabarti}(2008)}]{annealing}%
  \BibitemOpen
  \bibfield  {author} {\bibinfo {author} {\bibfnamefont {A.}~\bibnamefont
  {Das}}\ and\ \bibinfo {author} {\bibfnamefont {B.~K.}\ \bibnamefont
  {Chakrabarti}},\ }\href {\doibase 10.1103/RevModPhys.80.1061} {\bibfield
  {journal} {\bibinfo  {journal} {Rev. Mod. Phys.}\ }\textbf {\bibinfo {volume}
  {80}},\ \bibinfo {pages} {1061} (\bibinfo {year} {2008})}\BibitemShut
  {NoStop}%
\bibitem [{\citenamefont {Johnson}\ \emph {et~al.}(2011)\citenamefont
  {Johnson}, \citenamefont {Amin}, \citenamefont {Gildert}, \citenamefont
  {Lanting}, \citenamefont {Hamze}, \citenamefont {Dickson}, \citenamefont
  {Harris}, \citenamefont {Berkley}, \citenamefont {Johansson}, \citenamefont
  {Bunyk}, \citenamefont {Chapple}, \citenamefont {Enderud}, \citenamefont
  {Hilton}, \citenamefont {Karimi}, \citenamefont {Ladizinsky}, \citenamefont
  {Ladizinsky}, \citenamefont {Oh}, \citenamefont {Perminov}, \citenamefont
  {Rich}, \citenamefont {Thom}, \citenamefont {Tolkacheva}, \citenamefont
  {Truncik}, \citenamefont {Uchaikin}, \citenamefont {Wang}, \citenamefont
  {Wilson},\ and\ \citenamefont {Rose}}]{dwave}%
  \BibitemOpen
  \bibfield  {author} {\bibinfo {author} {\bibfnamefont {M.~W.}\ \bibnamefont
  {Johnson}}, \bibinfo {author} {\bibfnamefont {M.~H.~S.}\ \bibnamefont
  {Amin}}, \bibinfo {author} {\bibfnamefont {S.}~\bibnamefont {Gildert}},
  \bibinfo {author} {\bibfnamefont {T.}~\bibnamefont {Lanting}}, \bibinfo
  {author} {\bibfnamefont {F.}~\bibnamefont {Hamze}}, \bibinfo {author}
  {\bibfnamefont {N.}~\bibnamefont {Dickson}}, \bibinfo {author} {\bibfnamefont
  {R.}~\bibnamefont {Harris}}, \bibinfo {author} {\bibfnamefont {A.~J.}\
  \bibnamefont {Berkley}}, \bibinfo {author} {\bibfnamefont {J.}~\bibnamefont
  {Johansson}}, \bibinfo {author} {\bibfnamefont {P.}~\bibnamefont {Bunyk}},
  \bibinfo {author} {\bibfnamefont {E.~M.}\ \bibnamefont {Chapple}}, \bibinfo
  {author} {\bibfnamefont {C.}~\bibnamefont {Enderud}}, \bibinfo {author}
  {\bibfnamefont {J.~P.}\ \bibnamefont {Hilton}}, \bibinfo {author}
  {\bibfnamefont {K.}~\bibnamefont {Karimi}}, \bibinfo {author} {\bibfnamefont
  {E.}~\bibnamefont {Ladizinsky}}, \bibinfo {author} {\bibfnamefont
  {N.}~\bibnamefont {Ladizinsky}}, \bibinfo {author} {\bibfnamefont
  {T.}~\bibnamefont {Oh}}, \bibinfo {author} {\bibfnamefont {I.}~\bibnamefont
  {Perminov}}, \bibinfo {author} {\bibfnamefont {C.}~\bibnamefont {Rich}},
  \bibinfo {author} {\bibfnamefont {M.~C.}\ \bibnamefont {Thom}}, \bibinfo
  {author} {\bibfnamefont {E.}~\bibnamefont {Tolkacheva}}, \bibinfo {author}
  {\bibfnamefont {C.~J.~S.}\ \bibnamefont {Truncik}}, \bibinfo {author}
  {\bibfnamefont {S.}~\bibnamefont {Uchaikin}}, \bibinfo {author}
  {\bibfnamefont {J.}~\bibnamefont {Wang}}, \bibinfo {author} {\bibfnamefont
  {B.}~\bibnamefont {Wilson}}, \ and\ \bibinfo {author} {\bibfnamefont
  {G.}~\bibnamefont {Rose}},\ }\href {\doibase doi:10.1038/nature10012}
  {\bibfield  {journal} {\bibinfo  {journal} {Nature}\ }\textbf {\bibinfo
  {volume} {473}},\ \bibinfo {pages} {194} (\bibinfo {year}
  {2011})}\BibitemShut {NoStop}%
\bibitem [{\citenamefont {Boixo}\ \emph {et~al.}(2014)\citenamefont {Boixo},
  \citenamefont {Rønnow}, \citenamefont {Isakov}, \citenamefont {Wang},
  \citenamefont {Wecker}, \citenamefont {Lidar}, \citenamefont {Martinis},\
  and\ \citenamefont {Matthias}}]{troyer}%
  \BibitemOpen
  \bibfield  {author} {\bibinfo {author} {\bibfnamefont {S.}~\bibnamefont
  {Boixo}}, \bibinfo {author} {\bibfnamefont {T.~F.}\ \bibnamefont {Rønnow}},
  \bibinfo {author} {\bibfnamefont {S.~V.}\ \bibnamefont {Isakov}}, \bibinfo
  {author} {\bibfnamefont {Z.}~\bibnamefont {Wang}}, \bibinfo {author}
  {\bibfnamefont {D.}~\bibnamefont {Wecker}}, \bibinfo {author} {\bibfnamefont
  {D.~A.}\ \bibnamefont {Lidar}}, \bibinfo {author} {\bibfnamefont {J.~M.}\
  \bibnamefont {Martinis}}, \ and\ \bibinfo {author} {\bibfnamefont
  {T.}~\bibnamefont {Matthias}},\ }\href {\doibase 10.1038/nphys2900}
  {\bibfield  {journal} {\bibinfo  {journal} {Nature Physics}\ }\textbf
  {\bibinfo {volume} {10}},\ \bibinfo {pages} {218} (\bibinfo {year}
  {2014})}\BibitemShut {NoStop}%
\bibitem [{\citenamefont {Sjöqvist}\ \emph {et~al.}(2012)\citenamefont
  {Sjöqvist}, \citenamefont {Tong}, \citenamefont {Andersson}, \citenamefont
  {Hessmo}, \citenamefont {Johansson},\ and\ \citenamefont
  {Singh}}]{fastholonomic}%
  \BibitemOpen
  \bibfield  {author} {\bibinfo {author} {\bibfnamefont {E.}~\bibnamefont
  {Sjöqvist}}, \bibinfo {author} {\bibfnamefont {D.~M.}\ \bibnamefont {Tong}},
  \bibinfo {author} {\bibfnamefont {L.~M.}\ \bibnamefont {Andersson}}, \bibinfo
  {author} {\bibfnamefont {B.}~\bibnamefont {Hessmo}}, \bibinfo {author}
  {\bibfnamefont {M.}~\bibnamefont {Johansson}}, \ and\ \bibinfo {author}
  {\bibfnamefont {K.}~\bibnamefont {Singh}},\ }\href {\doibase
  doi:10.1088/1367-2630/14/10/103035} {\bibfield  {journal} {\bibinfo
  {journal} {New Journal of Physics}\ }\textbf {\bibinfo {volume} {14}},\
  \bibinfo {pages} {103035} (\bibinfo {year} {2012})}\BibitemShut {NoStop}%
\bibitem [{\citenamefont {Zanardi}\ and\ \citenamefont
  {Rasetti}(1999)}]{holonomic}%
  \BibitemOpen
  \bibfield  {author} {\bibinfo {author} {\bibfnamefont {P.}~\bibnamefont
  {Zanardi}}\ and\ \bibinfo {author} {\bibfnamefont {M.}~\bibnamefont
  {Rasetti}},\ }\href {\doibase 10.1016/S0375-9601(99)00803-8} {\bibfield
  {journal} {\bibinfo  {journal} {Physics Letters A}\ }\textbf {\bibinfo
  {volume} {264}},\ \bibinfo {pages} {94} (\bibinfo {year} {1999})}\BibitemShut
  {NoStop}%
\bibitem [{\citenamefont {Pachos}(2012)}]{topologicalQCbook}%
  \BibitemOpen
  \bibfield  {author} {\bibinfo {author} {\bibfnamefont {J.~K.}\ \bibnamefont
  {Pachos}},\ }\href@noop {} {\bibfield  {journal} {\bibinfo  {journal}
  {Introduction to Topological Quantum Computation, Cambridge University
  Press}\ } (\bibinfo {year} {2012})}\BibitemShut {NoStop}%
\bibitem [{\citenamefont {Bergmann}\ \emph {et~al.}(1998)\citenamefont
  {Bergmann}, \citenamefont {Theuer},\ and\ \citenamefont
  {Shore}}]{reviewSTIRAP}%
  \BibitemOpen
  \bibfield  {author} {\bibinfo {author} {\bibfnamefont {K.}~\bibnamefont
  {Bergmann}}, \bibinfo {author} {\bibfnamefont {H.}~\bibnamefont {Theuer}}, \
  and\ \bibinfo {author} {\bibfnamefont {B.~W.}\ \bibnamefont {Shore}},\ }\href
  {\doibase 10.1103/RevModPhys.70.1003} {\bibfield  {journal} {\bibinfo
  {journal} {Review of Modern Physics}\ }\textbf {\bibinfo {volume} {70}},\
  \bibinfo {pages} {1003} (\bibinfo {year} {1998})}\BibitemShut {NoStop}%
\bibitem [{\citenamefont {Vitanov}\ \emph {et~al.}(2001)\citenamefont
  {Vitanov}, \citenamefont {Halfmann}, \citenamefont {Shore},\ and\
  \citenamefont {Bergmann}}]{another_reviewSTIRAP}%
  \BibitemOpen
  \bibfield  {author} {\bibinfo {author} {\bibfnamefont {N.~V.}\ \bibnamefont
  {Vitanov}}, \bibinfo {author} {\bibfnamefont {T.}~\bibnamefont {Halfmann}},
  \bibinfo {author} {\bibfnamefont {B.~W.}\ \bibnamefont {Shore}}, \ and\
  \bibinfo {author} {\bibfnamefont {K.}~\bibnamefont {Bergmann}},\ }\href
  {\doibase 10.1146/annurev.physchem.52.1.763} {\bibfield  {journal} {\bibinfo
  {journal} {Annu. Rev. Phys. Chem}\ }\textbf {\bibinfo {volume} {52}},\
  \bibinfo {pages} {763} (\bibinfo {year} {2001})}\BibitemShut {NoStop}%
\bibitem [{\citenamefont {Gaubatz}\ \emph {et~al.}(1998)\citenamefont
  {Gaubatz}, \citenamefont {Rudecki}, \citenamefont {Schiemann},\ and\
  \citenamefont {Bergmann}}]{stirapfirst}%
  \BibitemOpen
  \bibfield  {author} {\bibinfo {author} {\bibfnamefont {U.}~\bibnamefont
  {Gaubatz}}, \bibinfo {author} {\bibfnamefont {P.}~\bibnamefont {Rudecki}},
  \bibinfo {author} {\bibfnamefont {S.}~\bibnamefont {Schiemann}}, \ and\
  \bibinfo {author} {\bibfnamefont {K.}~\bibnamefont {Bergmann}},\ }\href
  {\doibase 10.1063/1.458514} {\bibfield  {journal} {\bibinfo  {journal} {The
  Journal of Chemical Physics}\ }\textbf {\bibinfo {volume} {92}},\ \bibinfo
  {pages} {1990} (\bibinfo {year} {1998})}\BibitemShut {NoStop}%
\bibitem [{\citenamefont {Du}\ \emph {et~al.}(2014)\citenamefont {Du},
  \citenamefont {Liang}, \citenamefont {Huang}, \citenamefont {Yan},\ and\
  \citenamefont {Zhu}}]{experimental_atom_du}%
  \BibitemOpen
  \bibfield  {author} {\bibinfo {author} {\bibfnamefont {Y.-X.}\ \bibnamefont
  {Du}}, \bibinfo {author} {\bibfnamefont {Z.-T.}\ \bibnamefont {Liang}},
  \bibinfo {author} {\bibfnamefont {W.}~\bibnamefont {Huang}}, \bibinfo
  {author} {\bibfnamefont {H.}~\bibnamefont {Yan}}, \ and\ \bibinfo {author}
  {\bibfnamefont {S.-L.}\ \bibnamefont {Zhu}},\ }\href {\doibase
  10.1103/PhysRevA.90.023821} {\bibfield  {journal} {\bibinfo  {journal} {Phys.
  Rev. A}\ }\textbf {\bibinfo {volume} {90}},\ \bibinfo {pages} {023821}
  (\bibinfo {year} {2014})}\BibitemShut {NoStop}%
\bibitem [{\citenamefont {{Mangano, G.}}\ \emph {et~al.}(2008)\citenamefont
  {{Mangano, G.}}, \citenamefont {{Siewert, J.}},\ and\ \citenamefont {{Falci,
  G.}}}]{stirap_cooper_pair_box_falci}%
  \BibitemOpen
  \bibfield  {author} {\bibinfo {author} {\bibnamefont {{Mangano, G.}}},
  \bibinfo {author} {\bibnamefont {{Siewert, J.}}}, \ and\ \bibinfo {author}
  {\bibnamefont {{Falci, G.}}},\ }\href {\doibase 10.1140/epjst/e2008-00729-4}
  {\bibfield  {journal} {\bibinfo  {journal} {Eur. Phys. J. Special Topics}\
  }\textbf {\bibinfo {volume} {160}},\ \bibinfo {pages} {259} (\bibinfo {year}
  {2008})}\BibitemShut {NoStop}%
\bibitem [{\citenamefont {Falci}\ \emph {et~al.}(2013)\citenamefont {Falci},
  \citenamefont {La~Cognata}, \citenamefont {Berritta}, \citenamefont
  {D'Arrigo}, \citenamefont {Paladino},\ and\ \citenamefont
  {Spagnolo}}]{lambda_quantronium_falci}%
  \BibitemOpen
  \bibfield  {author} {\bibinfo {author} {\bibfnamefont {G.}~\bibnamefont
  {Falci}}, \bibinfo {author} {\bibfnamefont {A.}~\bibnamefont {La~Cognata}},
  \bibinfo {author} {\bibfnamefont {M.}~\bibnamefont {Berritta}}, \bibinfo
  {author} {\bibfnamefont {A.}~\bibnamefont {D'Arrigo}}, \bibinfo {author}
  {\bibfnamefont {E.}~\bibnamefont {Paladino}}, \ and\ \bibinfo {author}
  {\bibfnamefont {B.}~\bibnamefont {Spagnolo}},\ }\href {\doibase
  10.1103/PhysRevB.87.214515} {\bibfield  {journal} {\bibinfo  {journal} {Phys.
  Rev. B}\ }\textbf {\bibinfo {volume} {87}},\ \bibinfo {pages} {214515}
  (\bibinfo {year} {2013})}\BibitemShut {NoStop}%
\bibitem [{\citenamefont {Koch}\ \emph {et~al.}(2007)\citenamefont {Koch},
  \citenamefont {Yu}, \citenamefont {Gambetta}, \citenamefont {Houck},
  \citenamefont {Schuster}, \citenamefont {Majer}, \citenamefont {Blais},
  \citenamefont {Devoret}, \citenamefont {Girvin},\ and\ \citenamefont
  {Schoelkopf}}]{transmon_PRA2007}%
  \BibitemOpen
  \bibfield  {author} {\bibinfo {author} {\bibfnamefont {J.}~\bibnamefont
  {Koch}}, \bibinfo {author} {\bibfnamefont {T.~M.}\ \bibnamefont {Yu}},
  \bibinfo {author} {\bibfnamefont {J.}~\bibnamefont {Gambetta}}, \bibinfo
  {author} {\bibfnamefont {A.~A.}\ \bibnamefont {Houck}}, \bibinfo {author}
  {\bibfnamefont {D.~I.}\ \bibnamefont {Schuster}}, \bibinfo {author}
  {\bibfnamefont {J.}~\bibnamefont {Majer}}, \bibinfo {author} {\bibfnamefont
  {A.}~\bibnamefont {Blais}}, \bibinfo {author} {\bibfnamefont {M.~H.}\
  \bibnamefont {Devoret}}, \bibinfo {author} {\bibfnamefont {S.~M.}\
  \bibnamefont {Girvin}}, \ and\ \bibinfo {author} {\bibfnamefont {R.~J.}\
  \bibnamefont {Schoelkopf}},\ }\href {\doibase 10.1103/PhysRevA.76.042319}
  {\bibfield  {journal} {\bibinfo  {journal} {Phys. Rev. A}\ }\textbf {\bibinfo
  {volume} {76}},\ \bibinfo {pages} {042319} (\bibinfo {year}
  {2007})}\BibitemShut {NoStop}%
\bibitem [{\citenamefont {Wallraff}\ \emph {et~al.}(2004)\citenamefont
  {Wallraff}, \citenamefont {Schuster}, \citenamefont {Blais}, \citenamefont
  {Frunzio}, \citenamefont {Huang}, \citenamefont {Majer}, \citenamefont
  {Kumar}, \citenamefont {Girvin},\ and\ \citenamefont
  {Schoelkopf}}]{cqed_strong_coupling_wallraff}%
  \BibitemOpen
  \bibfield  {author} {\bibinfo {author} {\bibfnamefont {A.}~\bibnamefont
  {Wallraff}}, \bibinfo {author} {\bibfnamefont {D.~I.}\ \bibnamefont
  {Schuster}}, \bibinfo {author} {\bibfnamefont {A.}~\bibnamefont {Blais}},
  \bibinfo {author} {\bibfnamefont {L.}~\bibnamefont {Frunzio}}, \bibinfo
  {author} {\bibfnamefont {R.-S.}\ \bibnamefont {Huang}}, \bibinfo {author}
  {\bibfnamefont {J.}~\bibnamefont {Majer}}, \bibinfo {author} {\bibfnamefont
  {S.}~\bibnamefont {Kumar}}, \bibinfo {author} {\bibfnamefont {S.~M.}\
  \bibnamefont {Girvin}}, \ and\ \bibinfo {author} {\bibfnamefont {R.~J.}\
  \bibnamefont {Schoelkopf}},\ }\href {\doibase 10.1038/nature02851} {\bibfield
   {journal} {\bibinfo  {journal} {Nature}\ }\textbf {\bibinfo {volume}
  {431}},\ \bibinfo {pages} {162} (\bibinfo {year} {2004})}\BibitemShut
  {NoStop}%
\bibitem [{\citenamefont {Bianchetti}\ \emph {et~al.}(2009)\citenamefont
  {Bianchetti}, \citenamefont {Filipp}, \citenamefont {Baur}, \citenamefont
  {Fink}, \citenamefont {G\"oppl}, \citenamefont {Leek}, \citenamefont
  {Steffen}, \citenamefont {Blais},\ and\ \citenamefont
  {Wallraff}}]{dispersive_readout_bianchetti}%
  \BibitemOpen
  \bibfield  {author} {\bibinfo {author} {\bibfnamefont {R.}~\bibnamefont
  {Bianchetti}}, \bibinfo {author} {\bibfnamefont {S.}~\bibnamefont {Filipp}},
  \bibinfo {author} {\bibfnamefont {M.}~\bibnamefont {Baur}}, \bibinfo {author}
  {\bibfnamefont {J.~M.}\ \bibnamefont {Fink}}, \bibinfo {author}
  {\bibfnamefont {M.}~\bibnamefont {G\"oppl}}, \bibinfo {author} {\bibfnamefont
  {P.~J.}\ \bibnamefont {Leek}}, \bibinfo {author} {\bibfnamefont
  {L.}~\bibnamefont {Steffen}}, \bibinfo {author} {\bibfnamefont
  {A.}~\bibnamefont {Blais}}, \ and\ \bibinfo {author} {\bibfnamefont
  {A.}~\bibnamefont {Wallraff}},\ }\href {\doibase 10.1103/PhysRevA.80.043840}
  {\bibfield  {journal} {\bibinfo  {journal} {Phys. Rev. A}\ }\textbf {\bibinfo
  {volume} {80}},\ \bibinfo {pages} {043840} (\bibinfo {year}
  {2009})}\BibitemShut {NoStop}%
\bibitem [{\citenamefont {Vasilev}\ \emph {et~al.}(2009)\citenamefont
  {Vasilev}, \citenamefont {Kuhn},\ and\ \citenamefont
  {Vitanov}}]{optimal_stirap_pulses}%
  \BibitemOpen
  \bibfield  {author} {\bibinfo {author} {\bibfnamefont {G.~S.}\ \bibnamefont
  {Vasilev}}, \bibinfo {author} {\bibfnamefont {A.}~\bibnamefont {Kuhn}}, \
  and\ \bibinfo {author} {\bibfnamefont {N.~V.}\ \bibnamefont {Vitanov}},\
  }\href {\doibase 10.1103/PhysRevA.80.013417} {\bibfield  {journal} {\bibinfo
  {journal} {Phys. Rev. A}\ }\textbf {\bibinfo {volume} {80}},\ \bibinfo
  {pages} {013417} (\bibinfo {year} {2009})}\BibitemShut {NoStop}%
\bibitem [{\citenamefont {Torosov}\ and\ \citenamefont
  {Vitanov}(2013)}]{composite_stirap}%
  \BibitemOpen
  \bibfield  {author} {\bibinfo {author} {\bibfnamefont {B.~T.}\ \bibnamefont
  {Torosov}}\ and\ \bibinfo {author} {\bibfnamefont {N.~V.}\ \bibnamefont
  {Vitanov}},\ }\href {\doibase 10.1103/PhysRevA.87.043418} {\bibfield
  {journal} {\bibinfo  {journal} {Phys. Rev. A}\ }\textbf {\bibinfo {volume}
  {87}},\ \bibinfo {pages} {043418} (\bibinfo {year} {2013})}\BibitemShut
  {NoStop}%
\bibitem [{\citenamefont {Bianchetti}\ \emph {et~al.}(2010)\citenamefont
  {Bianchetti}, \citenamefont {Filipp}, \citenamefont {Baur}, \citenamefont
  {Fink}, \citenamefont {Lang}, \citenamefont {Steffen}, \citenamefont
  {Boissonneault}, \citenamefont {Blais},\ and\ \citenamefont
  {Wallraff}}]{three_level_tomography_bianchetti}%
  \BibitemOpen
  \bibfield  {author} {\bibinfo {author} {\bibfnamefont {R.}~\bibnamefont
  {Bianchetti}}, \bibinfo {author} {\bibfnamefont {S.}~\bibnamefont {Filipp}},
  \bibinfo {author} {\bibfnamefont {M.}~\bibnamefont {Baur}}, \bibinfo {author}
  {\bibfnamefont {J.~M.}\ \bibnamefont {Fink}}, \bibinfo {author}
  {\bibfnamefont {C.}~\bibnamefont {Lang}}, \bibinfo {author} {\bibfnamefont
  {L.}~\bibnamefont {Steffen}}, \bibinfo {author} {\bibfnamefont
  {M.}~\bibnamefont {Boissonneault}}, \bibinfo {author} {\bibfnamefont
  {A.}~\bibnamefont {Blais}}, \ and\ \bibinfo {author} {\bibfnamefont
  {A.}~\bibnamefont {Wallraff}},\ }\href {\doibase
  10.1103/PhysRevLett.105.223601} {\bibfield  {journal} {\bibinfo  {journal}
  {Phys. Rev. Lett.}\ }\textbf {\bibinfo {volume} {105}},\ \bibinfo {pages}
  {223601} (\bibinfo {year} {2010})}\BibitemShut {NoStop}%
\bibitem [{\citenamefont {Vitanov}\ \emph {et~al.}(1999)\citenamefont
  {Vitanov}, \citenamefont {Suominen},\ and\ \citenamefont
  {Shore}}]{fractional_stirap}%
  \BibitemOpen
  \bibfield  {author} {\bibinfo {author} {\bibfnamefont {N.~V.}\ \bibnamefont
  {Vitanov}}, \bibinfo {author} {\bibfnamefont {K.-A.}\ \bibnamefont
  {Suominen}}, \ and\ \bibinfo {author} {\bibfnamefont {B.~W.}\ \bibnamefont
  {Shore}},\ }\href {http://stacks.iop.org/0953-4075/32/i=18/a=312} {\bibfield
  {journal} {\bibinfo  {journal} {Journal of Physics B: Atomic, Molecular and
  Optical Physics}\ }\textbf {\bibinfo {volume} {32}},\ \bibinfo {pages} {4535}
  (\bibinfo {year} {1999})}\BibitemShut {NoStop}%
\bibitem [{\citenamefont {Møller}\ \emph {et~al.}(2007)\citenamefont
  {Møller}, \citenamefont {Madsen},\ and\ \citenamefont
  {Mølmer}}]{geometricphasegate}%
  \BibitemOpen
  \bibfield  {author} {\bibinfo {author} {\bibfnamefont {D.}~\bibnamefont
  {Møller}}, \bibinfo {author} {\bibfnamefont {L.~B.}\ \bibnamefont {Madsen}},
  \ and\ \bibinfo {author} {\bibfnamefont {K.}~\bibnamefont {Mølmer}},\ }\href
  {\doibase 10.1103/PhysRevA.75.062302} {\bibfield  {journal} {\bibinfo
  {journal} {Phys. Rev. A}\ }\textbf {\bibinfo {volume} {75}},\ \bibinfo
  {pages} {062302} (\bibinfo {year} {2007})}\BibitemShut {NoStop}%
\bibitem [{\citenamefont {Bal}\ \emph {et~al.}(2015)\citenamefont {Bal},
  \citenamefont {Ansari}, \citenamefont {Orgiazzi}, \citenamefont {Lutchyn},\
  and\ \citenamefont {Lupascu}}]{lupascu}%
  \BibitemOpen
  \bibfield  {author} {\bibinfo {author} {\bibfnamefont {M.}~\bibnamefont
  {Bal}}, \bibinfo {author} {\bibfnamefont {M.~H.}\ \bibnamefont {Ansari}},
  \bibinfo {author} {\bibfnamefont {J.-L.}\ \bibnamefont {Orgiazzi}}, \bibinfo
  {author} {\bibfnamefont {R.~M.}\ \bibnamefont {Lutchyn}}, \ and\ \bibinfo
  {author} {\bibfnamefont {A.}~\bibnamefont {Lupascu}},\ }\href {\doibase
  10.1103/PhysRevB.91.195434} {\bibfield  {journal} {\bibinfo  {journal} {Phys.
  Rev. B}\ }\textbf {\bibinfo {volume} {91}},\ \bibinfo {pages} {195434}
  (\bibinfo {year} {2015})}\BibitemShut {NoStop}%
\bibitem [{\citenamefont {Anisimov}\ \emph {et~al.}(2011)\citenamefont
  {Anisimov}, \citenamefont {Dowling},\ and\ \citenamefont
  {Sanders}}]{sanders}%
  \BibitemOpen
  \bibfield  {author} {\bibinfo {author} {\bibfnamefont {P.~M.}\ \bibnamefont
  {Anisimov}}, \bibinfo {author} {\bibfnamefont {P.}~\bibnamefont {Dowling},
  \bibfnamefont {Jonathan}}, \ and\ \bibinfo {author} {\bibfnamefont {B.~C.}\
  \bibnamefont {Sanders}},\ }\href {\doibase 10.1103/PhysRevLett.107.163604}
  {\bibfield  {journal} {\bibinfo  {journal} {Phys. Rev. Lett.}\ }\textbf
  {\bibinfo {volume} {107}},\ \bibinfo {pages} {163604} (\bibinfo {year}
  {2011})}\BibitemShut {NoStop}%
\end{thebibliography}%
%\cite{}

%\acknowledgments

\noindent{\bf Acknowledgements}\\
We are very grateful to Giuseppe Falci for comments about the paper.  This work used the cryogenic facilities of the Low Temperature Laboratory at Aalto University. We acknowledge financial support from FQXi, V\"aisal\"a Foundation, the Academy of Finland (project 263457), and the Center of Excellence ``Low Temperature Quantum Phenomena and Devices'' (project 250280).

\noindent{\bf Author Contributions}\\
K.S.K. measured the spectrum and contributed substantially to the experimental setup;
A.V. designed the tomography protocol and wrote the code for the simulations; S.D. fabricated the sample and took part in the measurements; G.S.P. initialized and supervised the project and contributed to the theoretical analysis. A.V. and G.S.P. wrote the manuscript with additional contributions from the other authors.

\noindent{\bf Additional information}\\
{\bf Supplementary Information} accompanies this paper. %at http://www.nature.com/naturecommunications.\\
{\bf Competing financial interests:} The authors declare no competing financial interests.

\newpage

{\large \bf Supplementary material}

\section{Sample and measurement setup}
\begin{figure}[h]
\centering
\includegraphics[width=1.0\columnwidth]{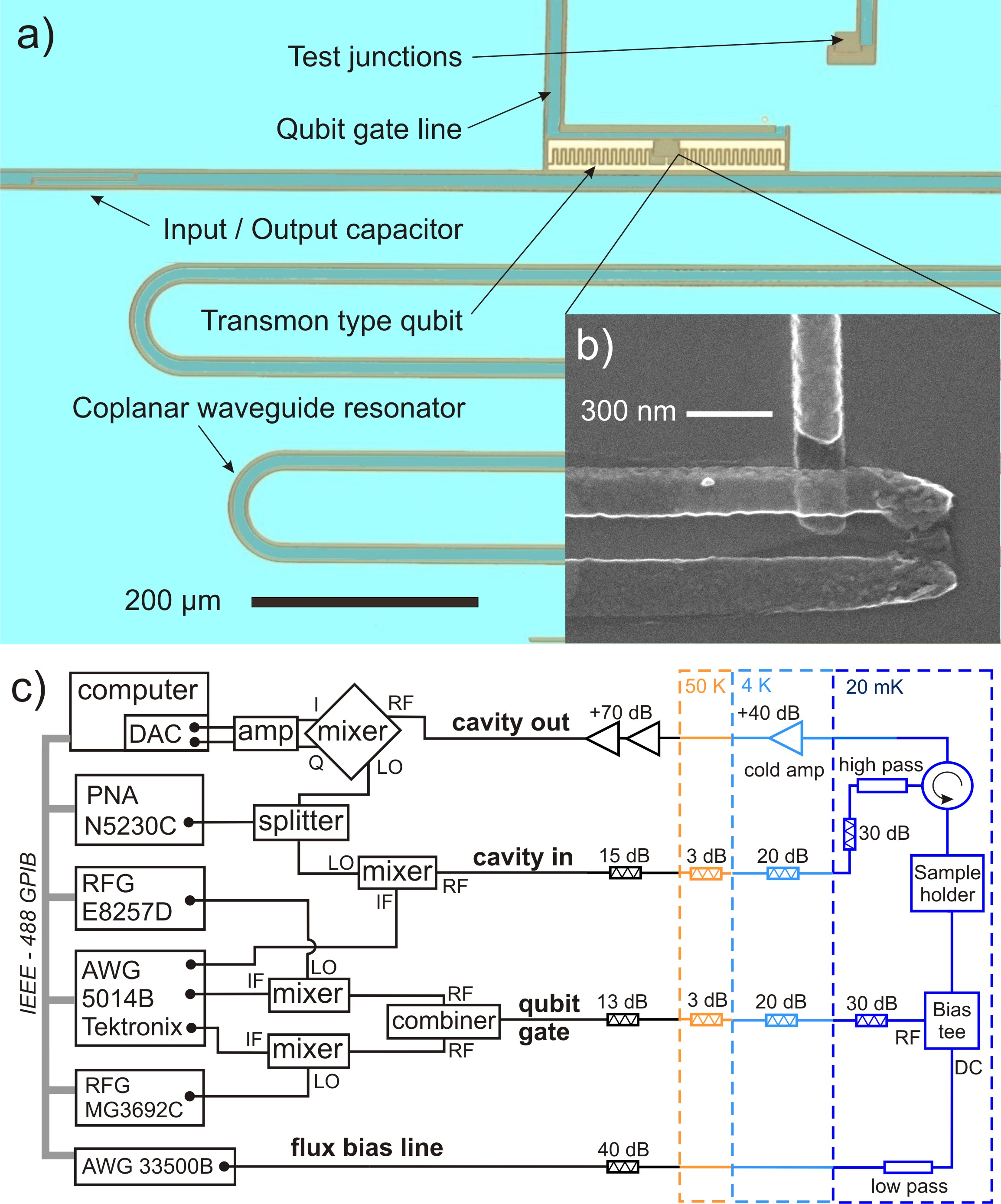}
\caption{a) Optical microscope image (false colors) of the main parts of the sample. b) SEM image of one of the Josephson junctions in the transmon. c) Simplified schematic of the room temperature electronics, and the cryogenic setup and wiring used in the experiment.}
\label{fig:sample_design}
\end{figure}

The superconducting QED circuit \cite{transmon_PRA2007} consists of a capacitively-shunted Cooper pair box (a transmon), with tunable energy level separation through the application of an external magnetic field. The qubit can be excited by applying a continuous wave or pulsed microwaves. The readout system is realized as a $\lambda /4$ coplanar waveguide resonator which allows the quantum non-demolition measurement of the state of the transmon \cite{cqed_strong_coupling_wallraff,dispersive_readout_bianchetti}. A picture of the sample is shown in Fig. \ref{fig:sample_design}a),b), and a schematic of the measurement setup at room temperature and in the dilution refrigerator is presented in Fig. \ref{fig:sample_design}c).

The transmon is essentially an artificial atom with energy levels $\hbar\omega_{j}$ yielding transition frequencies denoted by
$\omega_{j,j+1} = \omega_{j+1} - \omega_{j}$, and can be modelled as an anharmonic oscillator. With the standard notations in the field of superconducting devices, $E_{J}$ denoting the Josephson energy of the qubit and $E_{\rm C}$ the total charge energy (including the shunt capacitor), the anharmonicity is $\hbar \omega_{12}-\hbar \omega_{01} \approx - E_{\rm C}$ in the asymptotic limit $E_{\rm J} \gg E_{\rm C}$.
The usual way to manipulate the state of the circuit QED system is to apply  microwave drive signals
either to the gate of the transmon or to the coupled transmission line cavity. If the frequency of the drive $\omega_{j,j+1}^{(\Omega )}$
matches the transition frequency between two states of the system, the system goes through Rabi oscillations with frequency $\Omega_{j,j+1}$, resulting in transfer of population between the two states.

We model our system using the driven many-level Jaynes-Cummings Hamiltonian coupled to a resonator
of frequency $\omega_{\rm res}$. A measurement tone of frequency $\omega_{\rm meas}$ is coupled with strength $\epsilon_{\rm meas}$
into the resonator from the input capacitor of the coplanar waveguide cavity. The total Hamiltonian is then
%\begin{equation}
%\label{eq:jc}
%\begin{aligned}
%H_{JC} =& \frac{\hbar\omega_q}{2}\op{\sigma}_z + \hbar\omega_c(\op{a}^\dagger\op{a}+\frac{1}{2}) + \hbar g(\op{a}^\dagger\op{\sigma}_- +\op{a}\op{\sigma}_+) \\
%       +& \frac{\Omega_{0,1}}{2}(\op{\sigma}_-\ex{i\omega_{01}t}+ \op{\sigma}_+\ex{-i\omega_{01}t})\\
%       +& \frac{\Omega_{12}}{2}(\op{\sigma}_-\ex{i\omega_{12}t} + \op{\sigma}_+\ex{-i\omega_{12}t})
%\end{aligned}
%\end{equation}
\begin{equation}
\label{eq:jc_multilevel}
\begin{aligned}
\op{H}_{JC} &= \sum_{j=0}^{N}\hbar\omega_{j} \ket{j}\bra{j}
+ \hbar\omega_{\rm res}\op{a}^\dagger\op{a} \\
       &+ \sum_{j=0}^{N-1}\hbar g_{j,j+1}\left(\ket{j+1}\bra{j}\op{a} + \mathrm{h.c.} \right) \\
       &+ \sum_{j=0}^{N-1}\frac{\hbar\Omega_{j,j+1}}{2}\left(\ket{j+1}\bra{j}\ex{-i\omega_{j,j+1}^{(\Omega )}t} + \mathrm{h.c.}\right) \\
       &+ \hbar\epsilon_{\rm meas}\left(\op{a}^{\dag}\ex{-i\omega_{\rm meas}t} + \mathrm{h.c} \right),
\end{aligned}
\end{equation}
where $\ket{j}$ are the eigenstates of the $N$-level transmon corresponding to the eigenenergy $\hbar\omega_j$. Here we have applied the rotating wave approximation to the driving and the measurement fields, retaining only energy-conserving terms. We also only keep the coupling between consecutive energy levels with the corresponding near-resonant fields.

The single mode resonator, described by the annihilation (creation) operator $\op{a}$ ($\op{a}^\dagger$), couples to the qubit transition $j\rightarrow j+1$ with the Jaynes-Cummings coupling strength $g_{j,j+1}$.

In the dispersive regime, when the qubit is detuned from the cavity and the number of photons in the resonator is not too large such that
$4\langle \op{a}^{\dag} \op{a} \rangle
\left[g_{j,j+1}^2 /(\omega_{j+1} - \omega_j - \omega_{\rm res})^2\right] \ll 1$, we can
perform a generalized Schrieffer-Wolff transformation with displacement operator
$\op{D} = \exp \left[\op{A}\right]$, where
\begin{equation}
\op{A} = \sum_{j=0}^{N}\frac{g_{j,j+1}}{\omega_{j+1} - \omega_j - \omega_{\rm res}}\left(\op{a}\ket{j+1}\bra{j} - {\rm h. c.}\right),
\end{equation}
on the Hamiltonian in Eq. \eqref{eq:jc_multilevel}, resulting in $\op{H} \rightarrow \op{D}\op{H}\op{D}^{\dag}$.
Next, we use
\begin{equation}
e^{\op{A}}\op{H}e^{-\op{A}} = \op{H} + \left[\op{A}, \op{H}\right] + \frac{1}{2}\left[\op{A},\left[\op{A},\op{H}\right]\right],
\end{equation}
with $g_{j,j+1}/(\omega_{j+1} - \omega_j - \omega_{\rm res})$ as a small parameter and retaining only the first-order terms. This results in the cancellation of the Jaynes-Cummings interaction between the qubit and the resonator,
\begin{equation}
\label{eq:jc_dispersive_1}
\begin{aligned}
\op{H}_{\rm{disp-JC}} = &\sum_{j=1}^{N}\hbar\left(\omega_j  +
\chi_{j-1,j}\right)\ket{j}\bra{j} \\
              &-\hbar\chi_{0,1}\op{a}^\dagger\op{a}\ket{0}\bra{0}
              +\hbar \omega_{\rm res}\op{a}^\dagger\op{a}\\
              &+ \sum_{j=1}^{N} \hbar \left(\chi_{j-1,j}-\chi_{j,j+1}\right)\op{a}^\dagger\op{a}\ket{j}\bra{j} \\
              &+\sum_{j=0}^{N-1}\frac{\hbar\Omega_{j,j+1}}{2}\left(\ket{j+1}\bra{j}\ex{-i\omega_{j,j+1}^{(\Omega )}t} + \mathrm{h.c.}\right) \\
              &+ \hbar\epsilon_{\rm meas}\left(\op{a}^{\dag}\ex{-i\omega_{\rm meas}t} + \mathrm{h.c} \right).
\end{aligned}
\end{equation}
Here the dispersive shifts $\chi_{j,j+1}$ are defined as
\begin{equation}
\chi_{j,j+1} = \frac{g_{j,j+1}^2}{\omega_{j+1} - \omega_j - \omega_{\rm res}}.
\end{equation}
In obtaining Eq. (\ref{eq:jc_dispersive_1}), several terms have been neglected. A two-photon Hamiltonian resulting from the Jaynes-Cummings interaction couples states separated by two ladder indices,
\begin{equation}
\sum_{j=0}^{N-2}\frac{(2\omega_{j+1}-\omega_{j}-\omega_{j+2})\hbar g_{j,j+1}g_{j+1,j+2}}{2(
\omega_{j+2} - \omega_{j+1} - \omega_{\rm res})(\omega_{j+1} - \omega_{j} - \omega_{\rm res})}\op{a}^{2} |j+2\rangle \langle j| + {\rm h.c.}. \label{eq_twophoton_suppl}
\end{equation}
Then there are terms resulting from the commutator of $\hat{A}$ with the drive,
\begin{equation}
\begin{aligned}
&\sum_{j}\frac{g_{j,j+1}}{\omega_{j+1} - \omega_j - \omega_{\rm res}} a \left[
\frac{\hbar\Omega_{j-1,j}}{2}e^{-i\omega_{j-1,j}^{(\Omega )}t}|j+1\rangle\langle j-1| \right.\\
& -\frac{\hbar\Omega_{j+1,j+2}}{2}e^{-i\omega_{j+1,j+2}^{(\Omega )}t}|j+2\rangle\langle j|
+\frac{\hbar\Omega_{j,j+1}}{2}e^{i\omega_{j,j+1}^{(\Omega )}t} \\
& \left. \left(|j+1\rangle\langle j+1| - |j\rangle\langle j|\right)\right],
\end{aligned}
\label{eq_drivecorrection}
\end{equation}
and with the measurement pulse
\begin{equation}
\hbar \epsilon_{\rm meas} \sum_{j}\frac{g_{j,j+1}}{\omega_{j+1} - \omega_j - \omega_{\rm res}}\left[
|j+1\rangle\langle j|e^{-i\omega_{\rm meas} t} + {\rm h.c.}\right].
\end{equation}
Several arguments can be invoked to neglect these terms: some are second-order detuned processes such as Eq. (\ref{eq_twophoton_suppl}), and the rest would produce only first-order corrections to the dominant driving term. Moreover, when the qubit is driven, the resonator is not populated, and therefore the relevant terms left in Eq. (\ref{eq_drivecorrection}) contribute only as a renormalization of the decoherence of the second excited state into the resonator.

Next, we set the ground state as a reference for the energy levels by subtracting a quantity
$\hbar\omega_{0}\sum_{j=1}^{N}\ket{j}\bra{j}$ from the resulting Hamiltonian. Further, we move into a multiple-rotating frame defined by the transformation
\begin{equation}
\op{U}(t) = \sum_{j=1}^{N}\exp\left[i\sum_{k=0}^{j}\omega_{k,k+1}^{(\Omega)}t\right]\ket{j}\bra{j},
\end{equation}
which transforms the Hamiltonian as $\op{H} \rightarrow \op{U}\op{H}\op{U}^{\dag} + i\hbar(d \op{U}/dt)\op{U}^{\dag}$. We also perform a rotation with respect to the cavity at the measurement frequency, $\op{U}_{\rm meas} = \exp\left[ i \omega_{\rm meas } \op{a}^\dagger\op{a}t \right]$ resulting in a similar transformation $\op{H} \rightarrow \op{U}_{\rm meas}\op{H}\op{U}_{\rm meas}^{\dag} + i\hbar(d \op{U}_{\rm meas}/dt)\op{U}_{\rm meas}^{\dag}$. Finally, this yields
\begin{equation}
\label{eq:jc_dispersive}
\begin{aligned}
&\op{H}_{\rm{disp-JC}} = \sum_{j=1}^{N}\hbar\left(\omega_j - \omega_0 +
\chi_{j-1,j}
-\sum_{k=0}^{j}\omega_{k,k+1}^{(\Omega )}\right)\ket{j}\bra{j} \\
              &-\hbar\chi_{0,1}\op{a}^\dagger\op{a}\ket{0}\bra{0}
              +\hbar\left(\omega_{\rm res}-\omega_{\rm meas}\right)\op{a}^\dagger\op{a}\\
              &+ \sum_{j=1}^{N} \hbar \left(\chi_{j-1,j}-\chi_{j,j+1}\right)\op{a}^\dagger\op{a}\ket{j}\bra{j} \\
              &+ \sum_{j=0}^{N-1}\frac{\hbar\Omega_{j,j+1}}{2}\left(\ket{j+1}\bra{j} + \mathrm{h.c.}\right) + \hbar\epsilon_{\rm meas}\left(\op{a} + \op{a}^\dagger \right).
\end{aligned}
\end{equation}

Exciting the qubit on the energy level $j\geq 1$  results in a measurable shift of the resonator transition frequency by $\chi_{j-1,j}+\chi_{0,1}-\chi_{j,j+1}$, thus enabling a quantum non-demolition measurement of the transmon state. This is realized by monitoring the response of the cavity near the resonance under the application of a measurement pulse. Before the measurement field is applied, that is during the STIRAP pulses, the number of photons in the cavity, and therefore the ac-Stark shift on the transmon is negligible. In this case the only effect of the resonator on the qubit is that its vacuum fluctuations produce a small Lamb-shifted renormalization of the energy levels, $\tilde{\omega }_{j} = \omega_j +\chi_{j-1,j}$ for $j \geq 1$ and $\tilde{\omega }_{0} = \omega_0$, with corresponding transition frequencies $\tilde{\omega }_{j,j+1} = \tilde{\omega }_{j+1} -\tilde{\omega }_{j}$.
Let us introduce the detunings as
\begin{equation*}
\delta_{j,j+1} = \omega_{j+1} + \chi_{j,j+1} - \omega_j -\chi_{j-1,j}
- \omega_{j,j+1}^{(\Omega )}.
\end{equation*}
for $j \geq 1$ and $\delta_{01}=\omega_{1} + \chi_{01} - \omega_0 - \omega_{01}^{(\Omega )}$ for the first transition.

Next we focus on the STIRAP by considering only the three lowest eigenstates of the transmon, i.e. the states $\ket{0},\ket{1}$, and $\ket{2}$. This eventually leads to a very simple three-level form for the system Hamiltonian
% The transitions between these three states are performed by two drive signals, both having different tone. In the rotating wave approximation (RWA) the Hamiltonian of the system can be written
\begin{equation}
\label{eq:awg_hamiltonian_supplement}
\op{H}(t) = \frac{\hbar}{2}
\begin{bmatrix}
0 & \Omega_{01}(t) & 0 \\
\Omega_{01}(t) & 2\delta_{01} & \Omega_{12}(t) \\
0 & \Omega_{12}(t) & 2(\delta_{01} + \delta_{12})
\end{bmatrix}.
\end{equation}
To simplify the notation, in the expression above as well as in the main paper we have eliminated the comma between subscript indices whenever this does not lead to confusion, for example we write
$g_{0,1}\equiv g_{01}$, $\omega_{0,1}\equiv \omega_{01}$, $\Omega_{0,1}\equiv \Omega_{01}$.

\section{Derivation of the global adiabaticity condition}

Adiabatic processes should satisfy
$\abs{\bra{\pm}(d/dt)\ket{\rm D}}\ll \abs{\omega_\pm - \omega_{\rm D}}$,
%\left|\frac{\bra{\psi_m}\dot{\op{H}}\ket{\psi_n}}{E_n - E_m}\right| \ll 1, % Might not be correct. In other articles the condition is without H
which, when employing \eqref{eq:adiabatic_states} 
leads to
\begin{equation}
\label{eq:local_adiabatic_condition}
\abs{\frac{\dot{\Omega}_{01}(t)\Omega_{12}(t) - \Omega_{01}(t)\dot{\Omega}_{12}(t)}{\left[\Omega_{01}(t)^2 + \Omega_{12}(t)^2\right]^{3/2}}} \ll 1 ,
\end{equation}
for $\delta_{01} = 0$ and $\sin \Phi = 1$.

For Gaussian pulse shapes with equal amplitude
\begin{eqnarray}
\label{eq:gaussian_envelopes_S}
%\begin{aligned}
%\Omega_{01}(t) = \Omega\exp\left[-\frac{(t - t_{s}/2)^2}{2\sigma^2}\right], \\
%\Omega_{12}(t) = \Omega\exp\left[-\frac{(t + t_{s}/2)^2}{2\sigma^2}\right], \\
\Omega_{01}(t) &=& \Omega\exp\left[-\frac{t^2}{2\sigma^2}\right], \\
\Omega_{12}(t) &=& \Omega\exp\left[-\frac{(t - t_{s}/2)^2}{2\sigma^2}\right], 
%\end{aligned}
\end{eqnarray}
we can find a global adiabatic condition by integrating Eq. \eqref{eq:local_adiabatic_condition} over time
\begin{equation}
\frac{\pi}{2} \ll \dint{\sqrt{\Omega_{01}(t)^2 + \Omega_{12}(t)^2}}{t}{-\infty}{\infty} .
\end{equation}
The remaining integral does not have an analytical solution. However, it can be shown that it is an increasing function of $t_{s}$; moreover, the values at $t_{s}=0$ and $t_{s}=\infty$ can be calculated analytically, and they are $2\sqrt{\pi}\sigma\Omega$ and $2\sqrt{2\pi}\sigma\Omega$, respectively.
Thus, we have to select the minimum value $2\sqrt{\pi}\sigma\Omega$ to obtain the constraint, and we get
\begin{equation}
\label{eq:global_adiabatic_condition_supplement}
\frac{4}{\sqrt{\pi}}\sigma\Omega \gg 1.
\end{equation}

\section{Calibration and measurement protocol}

We first determine the transition frequencies of the transmon by measuring the Rabi frequency of a driven transition as a function of the drive frequency. This gives $\widetilde{\omega}_{01}/2\pi = $ 5.27 GHz, $\widetilde{\omega}_{12}/2\pi = $ 4.82 GHz.

The used measurement setup is shown in Fig. \ref{fig:sample_design}. To characterize the STIRAP process we determine the state of the three level transmon by probing the coupled transmission line cavity with a measurement signal. Due to the dispersive shift in the cavity resonance frequency, the response of the cavity to the probing microwave signal is different depending on the state of the transmon. To measure the cavity response we apply homodyne detection scheme, where the incoming microwave signal is downconverted to DC using an IQ-mixer. The resulting in-phase and quadrature signals are given by
\begin{equation}
\begin{aligned}
\avg{\op{I}(\tau)} &= -\eta\avg{\op{a}+\op{a}^\dagger} \\
\avg{\op{Q}(\tau)} &= i\eta\avg{\op{a}-\op{a}^\dagger},
\end{aligned}
\end{equation}
where $\eta$ is a factor describing the losses of the conversion and the other constants. By preparing the transmon in the states $\ket{0}$, $\ket{1}$, and $\ket{2}$ and then measuring the corresponding cavity response, we can associate the response of each of the states with a variable $r_j(\tau) = \{\avg{\op{I}_j(\tau)},\avg{\op{Q}_j(\tau)}\}$ and then describe the cavity response of every other state as a linear combination of $r_j(\tau)$
%\begin{equation}
%\begin{aligned}
%&r_m(t) = \alpha r_0(t) + \beta r_1(t) + \gamma r_2(t),& \\
%&\alpha + \beta + \gamma = 1,& 0 \le \alpha,\beta,\gamma \le 1. \\
%\end{aligned}
%\end{equation}
\begin{equation}
r_{\rm meas}(t,\tau) = \sum_{j=0,1,2} p_{j}(t) r_{j} (\tau ).
\label{eq_combi}
\end{equation}
This is a linear combination of the responses of each state, weighted by the population $p_{j}(t) = {\rm Tr} [\rho (t)\ket{j}\bra{j}]$ of the state, $\sum_{j=0,1,2} p_{j}(t)=1$ and $0 \le p_{j}(t) \le 1$ \cite{three_level_tomography_bianchetti}. This can be understood by introducing the operator corresponding to $r_{\rm meas }(t,\tau)$,
$\op{r}_{\rm meas} (\tau ) = \sum_{j=0,1,2} r_{j} (\tau )\ket{j}\bra{j}$
 and thus obtaining Eq. (\ref{eq_combi}) as $r_{\rm meas} (t,\tau) = {\rm Tr} [\rho (t)\op{r}_{\rm meas}(\tau )]$.

For a given trace $r_{\rm meas} (t,\tau)$ in the $I-Q$ plane, we determine $p_{0}$, $p_{1}$, and $p_{2}$ by applying the least square fit method to invert Eq. (\ref{eq_combi}). To implement the least squares fit method we employ the Levenberg-Marquardt algorithm, using the measurement response data up to $\tau = 1500$ ns. % using the measurement response data with $100 ns < \tau < 1500 ns$.
As pre-processing, the data is multiplied with an exponential weighting function $\exp [-\tau/w]$, with $w = 700$ ns. This allows us to use predominantly the data from the beginning of the
response curves, where the effect of decoherence on the distinguishability of the states is minimal.

%\begin{equation}
%\rho(t) = \alpha(t)\ket{0}\bra{0} + \beta(t)\ket{1}\bra{1} + \gamma(t)\ket{2}\bra{2}. \label{denstmatr}
%\end{equation}
In %Fig. \ref{fig:stirap_measurement}c)
Fig. 1c) we present the  $\avg{\op{I}(\tau)}$ traces corresponding to the system in
the ground state (blue), first excited state (green), and second excited state (red), together with the trace for the density matrix at a time $t=450$ ns (cyan) during the STIRAP protocol.
The calibration trace of the first excited state is obtained by applying a resonant $\pi$ pulse to the $\ket{0}\rightarrow\ket{1}$ transition, while the calibration trace of the second excited state is obtained by first populating the first excited state with a $\pi$ pulse, and then applying another $\pi$ pulse in resonance with the $\ket{1}\rightarrow\ket{2}$ transition. This allows us to infer the density matrix of the system, given a set of measured $\avg{\op{I}(\tau )}, \avg{\op{Q} (\tau )}$ traces. The main source of error in this procedure is produced by decoherence. The decay from the upper states during the calibration pulses results in a small inaccuracy, which produces a small overestimation of the populations in the tomography. We minimize this error as postprocessing by including the effect of decoherence in the calibration.

 %In the case of a two-level system the result of the measurement is given by

%\begin{equation}
%s = \int_0^t W(t)
%\end{equation}

\section{STIRAP as a route to holonomic quantum computing}

The phases of the driving fields, which we ignored so far, can lead to interesting new ways of qubit manipulation through the accumulation of geometric phases. If the phase factors are retained in the
driving fields, we get the following three-level form for the system Hamiltonian
% The transitions between these three states are performed by two drive signals, both having different tone. In the rotating wave approximation (RWA) the Hamiltonian of the system can be written
\begin{equation}
\label{eq:awg_hamiltonian_holonomic_supplement}
\op{H}(t) = \frac{\hbar}{2}
\begin{bmatrix}
0 & \Omega_{01}^{*} & 0 \\
\Omega_{01} & 2\delta_{01} & \Omega_{12} \\
0 & \Omega_{12}^{*} & 2(\delta_{01} + \delta_{12})
\end{bmatrix}.
\end{equation}
Let us denote by $\varphi $ the phase difference between the driving fields, therefore we can take $\Omega_{01} = |\Omega_{01}|$, $\Omega_{12}= |\Omega_{01}|\exp (-i\varphi )$. At two-photon resonance $\delta_{01} + \delta_{12} = 0$, the eigenvalues/eigenvectors problem at a time $t$ has the solution $\omega_+ = \delta_{01} + \sqrt{\delta_{01}^2 + |\Omega_{01}|^2 + |\Omega_{12}|^2}$, $\omega_- = \delta_{01} - \sqrt{\delta_{01}^2 + |\Omega_{01}|^2 + |\Omega_{12}|^2}$,
and $\omega_{\rm D} = 0$, with corresponding eigenvectors
%\begin{equation}
%\label{eq:awg_eigenvalues}
%\begin{aligned}
%\omega_+ &= \delta_{01} + \sqrt{\delta_{01}^2 + \Omega_{01}^2 + \Omega_{12}^2}, \\
%\omega_- &= \delta_{01} - \sqrt{\delta_{01}^2 + \Omega_{01}^2 + \Omega_{12}^2}, \\
%\omega_0 &= 0.
%\end{aligned}
%\end{equation}
%\begin{equation}
%\begin{aligned}
%\cos{\Theta} &= \frac{\Omega_{12}(t)}{\sqrt{\Omega_{01}(t)^2 + \Omega_{12}(t)^2}},\, \tan{\Theta} &= %\frac{\Omega_{01}(t)}{\Omega_{12}(t)} \\
%\sin{\Theta} &= \frac{\Omega_{01}(t)}{\sqrt{\Omega_{01}(t)^2 + \Omega_{12}(t)^2}}, \\
%\end{aligned}
%\end{equation}
\begin{equation}
\label{eq:awg_states}
\begin{aligned}
\ket{+} &= \sin{\Phi}\ket{B} + \cos{\Phi}\ket{1}, \\
\ket{-} &= \cos{\Phi}\ket{B} - \sin{\Phi}\ket{1}, \\
\ket{\rm D} &= \cos{\Theta}\ket{0} - \sin{\Theta}e^{i\varphi}\ket{2},
\end{aligned}
\end{equation}
where the bright state is $\ket{B} = \sin{\Theta}\ket{0} + \cos{\Theta}e^{i\varphi}\ket{2}$.
The angle $\Theta$ is defined by $\tan{\Theta} = |\Omega_{01}(t)|/|\Omega_{12}(t)|$ and parameterizing the rotation in the $\{|0\rangle, |2\rangle \}$ subspace, and $\Phi$ is defined in the same way, as an angle of a right triangle with vertices $\sqrt{(|\Omega_{01}|^2 + |\Omega_{12}|^2)/2}$ and $\sqrt{(|\Omega_{01}|^2 + |\Omega_{12}|^2 + \delta_{01}^2)/2} + \delta_{01}/\sqrt{2}$.

The Berry phase accumulated during a STIRAP sequence can be calculated from the standard Berry's connection formula
\begin{equation}
\gamma_{\rm Berry} = i \int_{R_{i}}^{R_{f}} \langle {\rm D} \vert \nabla_{\bf{R}}\vert{\rm D}\rangle d{\rm R},
\end{equation}
where the integral is taken along a contour in the $(\Theta ,\varphi)$ space,
\begin{equation}
{\bf R} = \left(\begin{array}{c} \Theta \\ \varphi \end{array} \right).
\end{equation}
For a general trajectory starting at an initial time $t_{i}$, this yields at time $t$ \cite{geometricphasegate}
\begin{equation}
\gamma_{\rm Berry} (t) = - \int_{\varphi ({t_i})}^{\varphi (t)} \sin^{2}\Theta d\varphi .
\end{equation}
Because the energy eigenvalue of the dark state is zero, from the  time $t_{i}$ to time $t$ the system would only pick up a geometrical phase,
\begin{equation}
\ket{{\rm D}(t_{i})} \rightarrow e^{i\gamma_{\rm Berry}(t)} \ket{{\rm D}(t)}.
\end{equation}
After a full STIRAP process the state then changes as
\begin{equation}
\ket{0} \rightarrow -e^{i\gamma_{\rm Berry} (t_{f}) + i\varphi (t_{f})}\ket{2} .
\end{equation}
In holonomic quantum computing, the geometric phase $\gamma_{\rm Berry}$ is used to construct single-qubit phase gates.

%
% ****** End of file apssamp.tex ******

\end{document}